\documentclass[aps,prd,groupedaddress,nofootinbib]{revtex4-1}

\usepackage{dsfont}
\usepackage{natbib}
\usepackage{amssymb}
\usepackage{graphicx}
\usepackage{amsmath}
\usepackage{url}
\usepackage{epstopdf}
\usepackage{color}
\usepackage{enumerate}
\usepackage[hidelinks]{hyperref}
\usepackage{array}

\newcommand{\bra}[1]{\langle #1|}
\newcommand{\ket}[1]{|#1\rangle}
\newcommand{\braket}[2]{\langle #1|#2\rangle}
\newcommand{\expect}[1]{\langle #1 \rangle}
\renewcommand{\H}{\mathcal{H}}
\newcommand{\PP}{P_{\phi}}
\newcommand{\pp}{p_{\phi}}
\newcommand{\sq}{\sqrt{q}}
\newcommand{\sh}{\sqrt{h}}
\newcommand{\ka}{\kappa}
\newcommand{\kb}{\bar{\kappa}}

\newcommand{\pphi}{\partial_\phi}

\newcommand{\ba}{\bar{a}}
\newcommand{\bt}{\bar{t}}

\newcommand{\tpp}{\widetilde{p_{\phi}}}
\newcommand{\tm}{\widetilde{m}}
\newcommand{\tphi}{\widetilde{\phi}}
\newcommand{\tl}{\widetilde{L}}
\newcommand{\lla}{l_{\Lambda}}
\newcommand{\kk}{\textbf{k}}

\newcommand{\nl}{\newline}
\def\p {\partial}                                                                       
\def\be {\begin{equation}}                                                                
\def\ee  {\end{equation}}                                                                
\def\bea {\begin{eqnarray}}                                           
\def\eea {\end{eqnarray}}                                               
\def\nn {\nonumber}

\begin{document}

\title{A non-perturbative Hamiltonian approach to the cosmological constant problem}

\author{Syed Moeez Hassan}
\email{shassan@unb.ca}
\affiliation{Department of Mathematics and Statistics, University of New Brunswick, Fredericton, NB, Canada E3B 5A3} 

\date{\today}

\begin{abstract}

It was recently suggested that the cosmological constant problem as viewed in a non-perturbative framework is intimately connected to the choice of time and a physical Hamiltonian. We develop this idea further by calculating the non-perturbative vacuum energy density as a function of the cosmological constant with multiple choices of time. We also include a spatial curvature of the universe and generalize this calculation beyond cosmology at a classical level. We show that vacuum energy density depends on the choice of time, and in almost all time gauges, is a non-linear function of the cosmological constant. This non-linear relation is a calculation for the vacuum energy density given some arbitrary value of the cosmological constant. Hence, in this non-perturbative framework, the cosmological constant problem does not arise. We also discuss why the conventional cosmological constant problem is not well-posed, and formulate and answer the question: ``Does vacuum gravitate?'' Finally, we give a derivation of reduction to quantum mechanics on a flat background from a non-perturbative gravity-matter theory.

\end{abstract}

\maketitle

\section{Introduction}

The cosmological constant problem has been described as one of the greatest crises in modern physics. On the one hand, the cosmological constant appears in General Relativity as a parameter and explains the current observed expansion of our Universe, whose value is close to zero. On the other hand, a calculation from Quantum Field Theory (QFT) shows the vacuum energy density to be very large. Then, assuming a connection between General Relativity and QFT, which equates vacuum energy density to the cosmological constant, leads to the problem that the observed value of the cosmological constant is off from its theoretical prediction by 120 orders of magnitude.

The conventional cosmological constant problem arises by assuming that quantum matter, in its vacuum state, will lead to a backreaction on the background geometry. From a non-perturbative perspective however, gravity and matter form a whole, and should both be quantized (instead of fixing a background apriori). In this non-perturbative framework, a `matter-vacuum' does not exist in general, therefore the question of backreaction of some matter-vacuum becomes ill-posed. What is relevant instead is a vacuum of the full gravity-matter system, which is computed using some given value of the cosmological constant. Therefore, from a non-perturbative perspective, the conventional problem does not arise.

In \cite{PhysRevLett.116.061302} it was argued that the assumed connection between General Relativity and QFT (on a fixed background) is suspect and we should really look at the cosmological constant problem in a full non-perturbative gravity-matter theory. The basic idea is that to define a vacuum energy density, a physical Hamiltonian is needed, which comes after identifying a time. Hence the notion of time, vacuum energy and the cosmological constant are intimately connected. There, they considered a spatially flat FRW universe with a massive scalar field, chose volume of the Universe as time, and quantized the resultant Hamiltonian using usual Schrodinger quantization.

In this paper, we extend the results of \cite{PhysRevLett.116.061302} to look at the effects of different choices of time (other than volume time) within the cosmological sector. We include a non-zero spatial curvature of the universe and also perform a Fermionic quantization of the resultant Hamiltonian. We also calculate the energy density in the full theory (without reducing to cosmology) at a classical level with three different choices of time. Finally, we show a new way of reducing to quantum mechanics on flat space from the full theory. We find that the generic conclusions in \cite{PhysRevLett.116.061302} still remain valid:  (i) The cosmological constant problem is consistently formulated only in a non-perturbative gravity-matter theory. (ii) Vacuum is sensibly defined after one has a physical Hamiltonian, which comes after choosing a particular time. Different choices of time lead to different physical Hamiltonians (and hence different quantum theories), which lead to different vacuum energy densities. (iii) Vacuum energy density is, in general, a non-linear function of the cosmological constant and time. (iv) In this non-perturbative framework, the vacuum energy density (of the full gravity-matter system) is \emph{calculated} using a \emph{given} value of the cosmological constant, hence, the cosmological constant problem does not arise.

This paper is organized as follows: In sections \ref{cc_prob}-\ref{vac_grav}, we explain the problem, and give the background and setup for our calculations. Sections \ref{classical_vac}-\ref{reduction} then present the detailed calculations. In Sec. \ref{cc_prob}, we look at six versions of the cosmological constant problem. In Sec. \ref{time_ham} we take a look at the problem of time in quantum gravity and define what is a physical Hamiltonian; the starting point for all our calculations. In Sec. \ref{what_vac} we take a close look at the meaning of vacuum in a non-perturbative theory of gravity and matter. In Sec. \ref{vac_grav} we formulate and answer the question: ``Does vacuum gravitate?'' In Sec. \ref{classical_vac}, we look at the relation between vacuum energy density and the cosmological constant at the classical level in three different time gauges: (i) scalar field time, (ii) York time and (iii) dust time. In Sec. \ref{cosmo}, we reduce to an FRW Universe from the full theory and to homogeneous fields on this background. Choosing volume of the universe as time, we explore the resulting quantum theory by extending the results of \cite{PhysRevLett.116.061302} to include a non-zero spatial curvature of the universe and also perform a Fermionic quantization of the Hamiltonian. Then, we consider three other time gauges: (i) scale factor time, (ii) scalar field time and (iii) dust time in the cosmological context and quantize the resultant physical Hamiltonians where possible. Vacuum energy density (as a function of $\Lambda$) is calculated in all of these gauges. In Sec. \ref{inhomo_cosmo} we look at the effect of including inhomogeneities in the homogeneous sector. We give a sketch of derivation for the full cosmological perturbations case, and explicitly calculate the vacuum energy density in scale factor time, for a simple model. In Sec. \ref{reduction}, we present a new way to reduce to quantum mechanics on flat space from our non-perturbative formalism. We present our conclusions in Sec. \ref{conclusions}, and give a summary and some discussion in Sec. \ref{discussion}. (We work in $G=c=\hbar = 1$ units.)

\section{What is the cosmological constant problem?} \label{cc_prob}

The conventional formulation of the cosmological constant problem is in the context of quantum fields on a fixed background \cite{RevModPhys.61.1}. There are different ways of formulating the problem and there are also some other types of ``cosmological constant problems''. In what follows, we briefly review six types of the problem. The starting point is Einstein Field Equations (EFEs),
\be
G_{\mu \nu} + \Lambda g_{\mu \nu} = 8 \pi T_{\mu \nu},
\ee
or, in a form where the cosmological constant appears as some sort of a stress-energy tensor,
\be
\label{EFE2}
G_{\mu \nu} = 8 \pi \Big( T_{\mu \nu} - \dfrac{\Lambda}{8 \pi} g_{\mu \nu} \Big).
\ee
It is also useful to mention here that the observed value of the cosmological constant is \cite{Ade:2015xua},
\be
\label{lambda_obs}
\Lambda_{obs} \sim 3 \times 10^{-122} l_{P}^{-2},
\ee
where $l_P$ is the Planck length. For reviews of this problem in various forms, see \cite{Rugh:2000ji,Martin:2012bt,RevModPhys.61.1,lrr-2001-1,Padilla:2015aaa,Burgess:2013ara,Padmanabhan:2014nca}.

\subsection{A classical problem: Shift symmetry} \label{shift_prob}

A first type of the cosmological constant problem is inherently classical without any reference to quantum fields. It goes as follows \cite{Padmanabhan:2014nca}.

The matter equations of motion are invariant under a constant shift of the Lagrangian: $\mathcal{L}(\Phi, g_{\mu \nu}) \rightarrow \mathcal{L}(\Phi, g_{\mu \nu}) + C$ (here $\Phi$ denotes all the matter degrees of freedom.) If we now couple this Lagrangian to General Relativity, we get,
\be
S = \int ~d^4x \sqrt{-g} \Big[ \Big( \mathcal{L}(\Phi, g_{\mu \nu}) + C \Big) + \dfrac{1}{16 \pi} R  \Big].
\ee
It is then clear that the matter equations of motion remain unchanged under this constant shift but the gravitational equations of motion change by the term $\sqrt{-g} C$. Hence a symmetry of the matter sector has been broken by coupling it to gravity. The above action can be re-written as
\be
S = \int ~d^4x \sqrt{-g} \Big[\mathcal{L}(\Phi, g_{\mu \nu}) + \dfrac{1}{16 \pi} \Big(  R-2 \Lambda \Big) \Big],
\ee
where we have defined $\Lambda = -8 \pi C$, making it clear that the constant shift can be interpreted as a cosmological constant.\footnote{One can also have a bare cosmological constant and a shift in the matter Lagrangian, but the physics will only depend on the sum of these two.} The problem here is that one can introduce an arbitrary shift in the matter Lagrangian (without affecting matter dynamics) which would then either introduce a cosmological constant (if a bare one is not already included), or it would change its numerical value. This leads to an infinite range of cosmological constants (by changing $C$), only one of which describes our universe.

\subsection{Another (pseudo) classical problem: Arbitrariness and fine-tuning}

Another version of the argument at the classical level goes as follows \cite{lrr-2001-1}. The energy-momentum tensor of a scalar field in an arbitrary potential is,
\be
T_{\mu \nu} = \dfrac{1}{2} \partial_\mu \phi \partial_\nu \phi + \dfrac{1}{2} \partial^\alpha \phi \partial_\alpha \phi g_{\mu \nu} - V(\phi) g_{\mu \nu}.
\ee
The lowest energy state (or vacuum) of this system will be when there is zero kinetic energy i.e. $\partial_\mu \phi = 0$, and the field is in its potential minimum $\phi = \phi_0$. This gives,
\be
T_{\mu \nu}^{vac} = -V(\phi_0) g_{\mu \nu}.
\ee
Comparing this with (\ref{EFE2}), the cosmological constant term looks like a vacuum energy term (This is the origin of the idea that the cosmological constant is the vacuum energy \cite{lrr-2001-1}.) Hence, we can associate the cosmological constant with this ground state (vacuum) energy as:
\be
V(\phi_0) = \dfrac{\Lambda}{8 \pi}.
\ee
The problem is stated as follows: The observed cosmological constant is immensely small (in Planck units).\footnote{This is where an input from quantum theory is required: namely the comparison in Planck units. Purely classically, the Planck scale is not relevant. This is the only quantum input we require in this sub-section, everything else is classical and no reference is made to quantum fields.} There is no apriori reason for the minimum of the potential to be so small. Therefore, why is it so small? This is a naturalness issue which is discussed further below.

This same argument can be turned into a fine tuning issue by saying that the cosmological constant appearing in EFEs is a bare cosmological constant ($\Lambda_0$), and it can have any value. The observed cosmological constant is then, the sum of the two contributions,
\be
\Lambda_{obs} = \Lambda_0 + 8 \pi V(\phi_0).
\ee
Now, once again, since $\Lambda_{obs}$ is so small, $\Lambda_0$ and $V(\phi_0)$ must be precisely fine-tuned so that they cancel out each other to 122 decimal places (in Planck units).

\subsection{The naive quantum problem}

We now look at the quantum problem. It is usually understood as a failure of theoretical prediction of the value of the cosmological constant coming from quantum field theory on a fixed background, versus the observed cosmological constant \cite{RevModPhys.61.1,lrr-2001-1}. A rough sketch of the argument goes as follows: a quantum field theory calculation on a \emph{(Ricci)-flat background} gives the vacuum energy density
\be
\label{huge_vac}
\expect{\hat{\rho}_{vac}} = \int_0^K \dfrac{d^3k}{(2 \pi)^3} k = \dfrac{K^4}{8 \pi^2},
\ee
where $K$ is some high energy cutoff scale.

In General Relativity, all energy gravitates (not just energy differences), leading to the conclusion that vacuum energy must gravitate as well. By making this statement, we have already \emph{assumed} a connection between classical general relativity and quantum field theory which is usually expressed as the semiclassical Einstein field equations
\be
\label{scefe}
G_{\mu \nu} + \Lambda g_{\mu \nu} = 8 \pi \bra{\psi} \widehat{T}_{\mu \nu} \ket{\psi},
\ee
where $\ket{\psi}$ is some suitable (but arbitrary) semiclassical state.

Evaluating these equations perturbatively around a flat (Minkowski) background $\eta_{\mu \nu}$, with a known vacuum state $\ket{0}$, a matter stress energy tensor $\widehat{T}_{\mu \nu}$ evaluated on this (Minkowski) background, and applying the above argument leads to the fact that at first order we have,
\be
\Lambda \eta_{\mu \nu} = 8 \pi \bra{0} \widehat{T}_{\mu \nu} \ket{0} = 8 \pi \expect{\hat{\rho}_{vac}} \eta_{\mu \nu} \sim K^4 \eta_{\mu \nu}.
\ee 
Assuming a Planck scale cutoff for $K$ leads to the cosmological constant problem since the observed value of $\Lambda $ (\ref{lambda_obs}) is vastly smaller than the predicted value,
\be
\label{huge_lambda}
\Lambda \sim 0.3 ~ l_P^{-2}.
\ee
This huge contradiction of 120 orders of magnitude arises due to calculating the same quantity in two different ways: one is through fitting observations of the universe (at the largest scales) to General Relativity (\ref{lambda_obs}), and the other is a calculation (at the smallest scales) from QFT on a fixed background (\ref{huge_lambda}).

\subsection{The refined quantum problem}

\begin{figure}
\begin{center}
\includegraphics[width=0.5\columnwidth, height = 0.15\textheight]{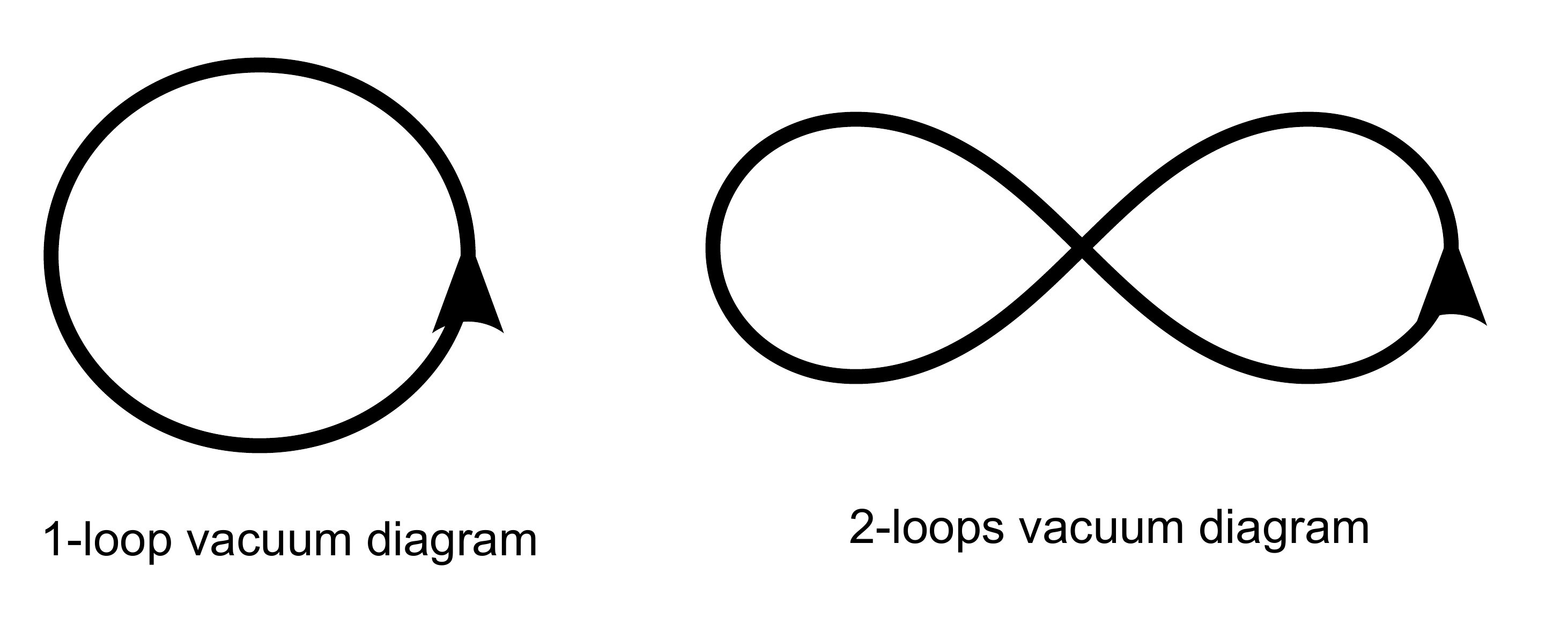}
\end{center}
\caption{Vacuum bubble diagrams at 1-loop and 2-loops.}\label{vacuum_bubbles}
\end{figure}

It has been argued that the above argument is misleading and the real problem is radiative corrections \cite{Padilla:2015aaa}. To make it concrete, consider a massive scalar field minimally coupled to classical gravity with a $\lambda \phi^4$ self interaction,
\be
\mathcal{L} =-\dfrac{1}{2} g^{\mu \nu} \p_\mu \phi \p_\nu \phi - \dfrac{1}{2} m^2 \phi^2 - \lambda \phi^4.
\ee
To calculate the vacuum energy in this formalism, we have to use perturbation theory (on some fixed background). The first contribution is at 1-loop (This is the scalar field loop with no external legs (figure \ref{vacuum_bubbles}). Recall that these are just the vacuum bubble diagrams which cancel out in the n-point functions in field theory with gravity turned off. With gravity however, they contribute to the cosmological constant term).

At 1-loop we get,
\be
V_{\mbox{vac}}^{{1-loop}} \sim -\dfrac{m^4}{(8 \pi)^2} \Bigg[\dfrac{2}{\epsilon} + \log \Bigg(\dfrac{\mu^2}{m^2} \Bigg) + \text{finite} \Bigg].
\ee
To cancel off this divergent contribution, one has to add to the bare cosmological constant a counterterm given as,
\be
\Lambda^{{1-loop}} \sim \dfrac{m^4}{(8 \pi)^2} \Bigg[\dfrac{2}{\epsilon} + \log \Bigg(\dfrac{\mu^2}{M^2} \Bigg)  \Bigg],
\ee
where $M$ is some arbitrary subtraction scale. This means that the value of the renormalized cosmological constant cannot be predicted, but has to be measured. This is just the usual field theoretic renormalization procedure, we cancel off the vacuum energy divergence with a suitable counterterm and are left with something finite.

The problem begins when we go to 2-loops. This is the eight figure with no external legs (figure \ref{vacuum_bubbles}). We get a contribution that goes as $\lambda m^4$. This is the same order of magnitude as the divergent term at 1-loop (For the Higgs boson for example, $\lambda \sim 0.1$). Therefore, we need to readjust the counterterm we added before at 1-loop to account for this as well. This is a retuning. But we can continue this to higher orders: 3-loops, 4-loops, and so on, requiring retuning at each successive order. And at each order, we have to fine-tune it to extreme accuracy. This is radiative instability and means that the cosmological constant is highly sensitive to high energy physics (physics which we do not know). This is another version of the problem.

A seemingly simple solution would be to sum the loops to all orders and then add the counter term. But the details of the loops are theory dependent and it is almost impossible, perturbatively at least, to do this to all orders.\footnote{See \cite{0264-9381-31-12-125006} for a case where an all loop analysis is done for the Gross-Neveu model. Note that it still does not include gravity as fully dynamical since they use the framework of semiclassical EFEs.} It seems that what is needed is a non-perturbative analysis of the situation, in the full gravity-matter theory.

\subsection{An effective argument}

Another way of looking at the cosmological constant problem is through the effective action \cite{Padilla:2015aaa,Burgess:2013ara}. It goes as follows: To describe the physics below a certain energy scale $\mu$, we take some ``fundamental'' (microscopic) action and integrate out the modes with energies higher than $\mu$. The basic point is that physics depends on the energy scale at which you look at the system.

Consider then a field theory in which we split the field (in Fourier space) into high energy and low energy modes,
\be
\phi = \phi_l + \phi_h,
\ee
where the modes $\phi_h$ lie above the energy scale $\mu$. The theory at energy scales below $\mu$ is then given by an effective action where we have integrated out $\phi_h$,
\be
\exp(i S_{\mbox{eff}}[\phi_l]) = \int D\phi_h \exp(i S[\phi_h,\phi_l]),
\ee
$S[\phi_h,\phi_l]$ is the microscopic action and $S_{\mbox{eff}}[\phi_l]$ is the action valid at scales below $\mu$.

Now, if we compute the vacuum energy using this effective action (it is important to note here that this is a calculation done on a fixed background), we expect it to be of the order of $\mu^4$, which can be large. In order to cancel this large contribution to get the actual (small) observed value of the cosmological constant, we need to add a cosmological counter term of the order of $\mu^4$.

But the mass scale $\mu$ is quite arbitrary. What happens if we integrate out a few more modes to get an effective action at a lower mass scale $\widetilde{\mu} < \mu$? We expect that any physics should not depend on the mass scale since it is just an arbitrary parameter. However, by the same argument as above, we expect the vacuum energy to be of the order of $\widetilde{\mu}^4$, so to cancel that, we have to add a cosmological counterterm of the order of $\widetilde{\mu}^4$. But previously, we already fixed this counter term to be of order $\mu^4$. We have to change it again i.e we have to retune it. But we can repeat the above argument again and again for any arbitrary mass scale, implying that we would have to retune this counter term again and again. This is the effective description of the problem. The tuning should not be sensitive to what low energy action we choose to do physics with.

\subsection{Naturalness issues} \label{natissue}

Underlying some of the above versions of the problem is another issue known as naturalness. This is not a seperate type of a cosmological constant problem, rather an additional question. The question is: why is the observed value of the cosmological constant so small compared to other scales that we know of. For an interesting solution (where it is suggested that a new physical principle is needed to answer this question) see \cite{Padmanabhan:2014nca,Padmanabhan:2016eld}.

\subsection{What will we address?}

Questions are posed and answered in a framework. The central feature of all the quantum versions of the cosmological constant problem stated above is the framework in which they are posed: quantum field theory on a fixed background, coupled in some way to general relativity (usually through the semiclassical Einstein Field Equations). Our point of view in this paper is that this is not the appropriate framework to address the cosmological constant problem. To quote Earman \cite{10.2307/41134106},

\begin{center}
\textit{``Rather than concluding that there is a``cosmological constant problem,'' one might alternatively conclude that there is something suspect either in the very the notion of vacuum energy density or else in the notion that this energy can serve as a source for the gravitational field.''}
\end{center}

Following \cite{PhysRevLett.116.061302}, we think that ``there is something suspect in the very notion of vacuum energy density'', as it is usually understood in QFT on a fixed background.\footnote{There are discussions on the other part of the quote above as well, that whether vacuum energy can serve as a source for gravity. See for example \cite{Jaffe:2005vp} for a discussion on the `reality' of zero point energy as established through the Casimir effect.} Our approach here is to start from a fully non-perturbative gravity-matter theory (note that this is a \emph{different} framework from QFT on flat space) and see if the cosmological constant problem (or a related one) arises there. We will not address the naturalness issue in this work with the view that $\Lambda$ should be determined by observations, as constants in a theory usually are.\footnote{For a view on naturalness issues that we agree with, see \cite{Hossenfelder:2018ikr}.} We now proceed to define our framework in the next section.

\section{The problem of time and a physical Hamiltonian} \label{time_ham}

Our starting point is the ADM action for general relativity minimally coupled to a massive scalar field,
\be
\label{start_action}
S = \int d^3x dt [ \pi^{ab} \dot{q}_{ab} + \widetilde{\PP} \dot{\phi} - N(\H_G + \H_{\phi}) - N^a (C^G_a + C^{\phi}_a)],
\ee
where,
\bea
\label{start_ham_full}
\H_G &=& \dfrac{1}{\sq} \Bigg(\pi^{ab} \pi_{ab} - \frac{1}{2} \pi^2 \Bigg) + \sq ( \Lambda - R), \nn \\
\H_{\phi} &=& \dfrac{\widetilde{\PP}^2}{2 \sq} + \dfrac{1}{2} \sq q^{ab} \partial_a \phi \partial_b \phi + \dfrac{1}{2} \sq m^2 \phi^2, \label{phi_ham} \\
C^{G}_a &=& D_b \pi^b_a, \nn \\
C^{\phi}_a &=& \widetilde{\PP} \partial_a \phi, \label{start_ham}
\eea
$(\pi^{ab},q_{ab})$ and $(\phi, \widetilde{\PP})$ are the gravitational and scalar field phase space variables respectively. $q$ is the determinant of the spatial metric $q_{ab}$, $\pi$ is the trace of $\pi^{ab}$, $R$ is the 3-Ricci scalar curvature, $\Lambda$ is the cosmological constant, $N$ is the shift, $N^a$ is the lapse and $m$ is the scalar field mass.

General relativity is a constrained theory, i.e. the Hamiltonian vanishes: $ \H_G=0 $. This constraint just reflects the time reparametrization invariance of general relativity.\footnote{Along with the diffeomorphism constraint which shows space reparametrization invariance, these four constraint equations are manifestations of the full diffeomorphism invariance of GR.} This leads to the (yet unsolved) problem of time in quantum gravity \cite{Isham:1992ms}. We remark on two ways to proceed from here to a quantum theory:

1. Start with the Hamiltonian constraint of GR and impose that the action of its corresponding quantum operator on all states in the Hilbert space must be zero. This leads to the Wheeler-deWitt equation:
\be
\widehat{H}_G \ket{\Psi} = 0,
\ee
with the matter Hamiltonian included as well, this becomes,
\be
\label{WDV}
\Big( \widehat{H}_G  + \widehat{H}_M \Big) \ket{\Psi} = 0.
\ee
2. Do a time gauge fixing (classically) to get a physical Hamiltonian.\footnote{We will call this the physical Hamiltonian approach. It is also known as the reduced phase space approach; and deparametrization.} This proceeds as follows:

\begin{itemize}

\item Identify a time function from the full phase space variables:
\be
\label{time_choice}
t = f(\mbox{phase space variables}).
\ee
Since there is no preferred notion of time, this is quite arbitrary and the only requirement is that (in Dirac's terminology) it should be second class with the Hamiltonian constraint: $\{t,H\} \not \approx 0$.

\item Let the momentum conjugate to this time function be $p_t$: $ \{ t,p_t\} = 1 $.

\item Solve the Hamiltonian constraint strongly for $p_t$. This is the \emph{non-vanishing} physical Hamiltonian corresponding to the specific time choice (\ref{time_choice}),
\be
H_p = -p_t.
\ee
\item Note that solving the Hamiltonian constraint strongly means that we have now gone from the full 14-dimensional phase space (12-gravity and 2-scalar field) to a partially reduced 12-dimensional physical phase space.\footnote{It is partial since we have not solved the diffeomorphism constraint. Solving it would further reduce the phase space to $12-(2 \times 3)=6$ dimensions: 4-gravity and 2-scalar field, which in configuration space corresponds to the usual 2 degrees of freedom for gravity, and 1 for the scalar field.} Two of the (phase space) degrees of freedom have vanished since they are now our time function and the physical Hamiltonian.

\item The requirement that the chosen gauge is preserved in time fixes the lapse:
\be
\dot{t} = 1 \Rightarrow \{t,H\} = \Big\{t,\int d^3x ~ N \H\Big\} = 1.
\ee
\end{itemize}

Once we have a physical Hamiltonian, we can go ahead and quantize it. \emph{This is the approach that we will follow in this paper.} Note that different choices of time will lead to different physical Hamiltonians and it is not clear whether the resulting quantum theories are equivalent in any sense.\footnote{ \label{ftn1} For instance, in \cite{PhysRevD.34.2323} we find the statement: \textit{``A different choice of time, Hilbert space, unitarily in-equivalent commutation relations, or operator ordering, for example, will lead to different canonical quantum theories.''}\\In the context of Loop Quantum Cosmology, it was shown in \cite{0264-9381-26-3-035012}, that different choices of lapse can lead to significantly different physical Hilbert spaces. For studies on how some physical predictions change by choosing a different time variable, see \cite{0264-9381-32-13-135004, 0264-9381-34-20-205001, 0264-9381-34-14-145012}.} 

A comment on the meaning of this physical Hamiltonian framework: within the framework of general relativity plus matter (where coordinate time has no physical relevance), we are identifying some quantities (phase space variables) that can serve as physical clocks. So suppose we have a scalar field $\phi$, and the spatial metric $q_{ab}$. Einstein's equations give the evolution of these quantities with respect to coordinate time. In this time-gauge fixed formalism however, we ask a different question: How do one of these quantities evolve with respect to the other one. For instance if we choose scalar field as time, then we ask how do all the other quantities like the spatial metric $q_{ab}$, or any other matter fields, evolve with respect to $\phi$. That is what is the value $q_{ab}$ takes, when $\phi=1,2,10$ etc. This is by construction, a gauge-invariant question, and has a well defined physical meaning. If we were to choose a different phase space variable as time, then the evolution of the remaining quantities, would in general look different (see also footnote \ref{ftn1}). Given that there is no preferred phase space variable to be used as time, this is just the (yet unsolved) problem of time. But each of the time choices that we make constitutes a well defined physical question: how do the rest of the phase space variables (in particular, the vacuum energy density) change in relation to this particular variable that we identify as time.

Now that our framework is defined, we will apply it to different choices of time and find the corresponding physical Hamiltonian. We can then calculate the vacuum energy density with each of those Hamiltonians. But before that, we need to define what we mean by a `vacuum'. This also leads to the question: `Does vacuum gravitate?'. We proceed to define a `vacuum' in the next section, and will return to formulate and answer this question in the subsequent section.

\section{What is a vacuum?} \label{what_vac}

This section will provide an answer to the question `what is a vacuum?' within the context of quantum theory, with the purpose in mind to formulate and answer the question `does vacuum gravitate?'. Since we want to treat the gravity-matter system as a whole, we will consider two \emph{different} vacua: vacuum of the full gravity-matter system (called `vacuum' from now on), and vacuum of the matter sub-system within the gravity-matter system (called `matter-vacuum'). We will first formally define both of these vacua, and then, will see if they are possible to define for two different approaches: The Wheeler-deWitt approach, and the Physical Hamiltonian approach, outlined in the previous section.

For any quantum system, the standard definition of vacuum comes after identifying a Hamiltonian $\widehat{H}$ of that particular system. So the first step is to find out (or write down) what is the Hamiltonian. Once this is done, the quantum vacuum state $\ket{0}$ of that system is most naturally \emph{defined} as the lowest energy eigenstate of that particular Hamiltonian,
\be
\label{full_vac}
\widehat{H} \ket{0} = E_0 \ket{0},
\ee
where $E_0$ is the lowest possible eigenvalue. This is what is meant by a vacuum in usual quantum theory.\footnote{It is important to note here that this is the vacuum of the \emph{full} physical system $\hat{H}$, and \emph{not} the vacuum of any particular subsystem (e.g, matter) within that Hamiltonian.} Given a gravity-matter system with a Hamiltonian $ \widehat{H}(\hat{g}, \hat{\Phi}) $ (where $\hat{g}$ denotes all the gravitational degrees of freedom, and $\hat{\Phi}$ all the matter ones), its quantum vacuum state will then be given by (\ref{full_vac}) as,
\be
\label{fullvac}
\widehat{H}(\hat{g},\hat{\Phi}) \ket{0} = E_0 \ket{0}.
\ee

Now, let us proceed to formally define a matter-vacuum. Again, before we can define a matter-vacuum, we must know what is the matter Hamiltonian $\widehat{H}_M$. There are two problems in proceeding further:

(i) Within the full non-perturbative theory of gravity and matter, the matter part exists as a subsystem, and in general, it may be that this subsystem is so mixed up with the gravity part, that it is not even possible to cleanly identify what is the matter part and what is the gravity part. We will see below that this is indeed the case in general for the physical Hamiltonian approach.

(ii) Since matter is coupled to gravity, the matter Hamiltonian $\widehat{H}_M$ will depend on the background metric $g$. This raises the question of whether we want to keep the background dynamical (i.e quantize the background: $\hat{g}$); or do we want to keep the background fixed (i.e have $g$), which is the arena for usual QFT on a fixed background.\footnote{If we quantize the background $\hat{g}$, then we have to solve the Hilbert space problem: A Hilbert space requires some notion of a metric (to compute inner products etc). Within the formalism of QFT on a fixed background, this metric is usually taken to be the background spacetime metric. But this is no longer possible if the background is also quantized.} From the fully non-pertubative approach in which both gravity and matter are dynamical, there is no apriori justification of fixing a background, and it must be shown that there is some limit of the theory in which using a fixed background makes sense. It is useful here to state a fundamental fact about General Relativity (and its quantum theory) \cite{Rovelli:2003wd}:

\begin{center}
\textit{``Physics on a curved spacetime is not GR. GR is the dynamics of spacetime itself. So, quantum GR is the theory of a quantum spacetime, not a quantum theory on various spacetimes.''}
\end{center}

Suppose, for the sake of argument, that somehow we solve problem (i) above, then the matter-vacuum can be defined as the lowest energy eigenstate of the matter Hamiltonian. But looking at (ii), we see that there will be two possibilities, one where the background is dynamical (hat on $g$),
\be
\label{matvacdyn}
\widehat{H}_M(\hat{g},\hat{\Phi}) \ket{0_M} = e_0 \ket{0_M},
\ee
and one where it is fixed (no hat on $g$),
\be
\label{matvacfix}
\widehat{H}_M(g,\hat{\Phi}) \ket{\tilde{0}_M} = \tilde{e}_0 \ket{\tilde{0}_M},
\ee
which leads to \emph{different} vacua: $\ket{0_M}$ lives in the full gravity-matter Hilbert space $\mathds{H}_{QG}$, whereas $\ket{\tilde{0}_M}$ lives in the matter Hilbert space $\mathds{H}_{M}$.

Now that we have the definition of a matter-vacuum above, and a vacuum as in (\ref{fullvac}), we see if they can actually be defined for General Relativity with matter for two different approaches.

\subsection{Vacuum and matter-vacuum in the Wheeler-deWitt approach}

Within the Wheeler-deWitt approach, we have the operator constraint (\ref{WDV}),
\be
\Big( \widehat{H}_G  + \widehat{H}_M \Big) \ket{\Psi} = 0. \nn
\ee

This means that \emph{all} physical states in the Hilbert space must satisfy this constraint, and hence \emph{all} states are eigenstates of the full Hamiltonian with the eigenvalue 0. This leads to an ambiguity in the notion of vacuum: all physical states have the lowest energy of 0, and therefore all of them are vacua. Clearly this notion of vacuum is not a very useful one.

Now, let us try to find a matter-vacuum in this approach. Since here, we have the notion of a matter Hamiltonian, we can proceed to define the matter-vacuum, with gravity as a dynamical variable as in (\ref{matvacdyn}),
\be
\widehat{H}_M(\hat{g},\hat{\Phi}) \ket{0_M} = e_0 \ket{0_M}. \nn
\ee
But this matter-vacuum state must also satisfy the Wheeler-deWitt equation,
\bea
\label{matgravvacwdw}
\Big( \widehat{H}_G  + \widehat{H}_M \Big) \ket{0_M} = 0 \nn \\
\Rightarrow \widehat{H}_G \ket{0_M} = - \widehat{H}_M \ket{0_M} = -e_0 \ket{0_M},
\eea

and hence, the matter-vacuum is also an eigenstate of the gravity sector. This however leads to,
\be
[\widehat{H}_M(\hat{g},\hat{\Phi}) , \widehat{H}_G(\hat{g}) ] \ket{0_M} = 0
\ee
and in general, we do not expect the matter and the gravitational Hamiltonians to commute.\footnote{This is because the matter Hamiltonian depends on the background on which it is defined and hence contains $\hat{g}$ in it. The gravity Hamiltonian contains the variables $\hat{\pi}$ which are canonically conjugate to $\hat{g}$: $[\hat{g},\hat{\pi}] \neq 0$.}

Note that there still can be states for which the above equation holds since we are not implementing a representation of the commutator of $\widehat{H}_M(\hat{g},\hat{\Phi})$ and $\widehat{H}_G(\hat{g})$, since they are not the fundamental variables (those being $\hat{g},\hat{\Phi}$ and their conjugates). But if there are such states, they will be highly special (in the sense that they would be eigenstates of both the matter and the gravity sector). We are not aware of any such states in the literature. In the absence of that, we can say that within the Wheeler-deWitt formalism, defining a matter-vacuum is highly non-trivial. Now let us turn to the physical Hamiltonian approach, the formalism we use in this paper.

\subsection{Vacuum and matter-vacuum in the Physical Hamiltonian approach}

Within the physical Hamiltonian approach, we arrive at a \emph{non-vanishing} physical Hamiltonian after fixing a time gauge. This means that we can compute its spectrum, and find the lowest eigenvalue state. Therefore, in contrast to the Wheeler-deWitt approach, it is possible here to find (in principle) the vacuum state of the full gravity-matter system. This will be the approach we follow in this paper, we will consider the full vacuum of the theory (and not just a matter-vacuum), obtained after fixing a time gauge and getting a physical Hamiltonian.

To complete the argument being made here, and to setup for the next section, let us now consider a matter-vacuum in this approach. As we will see later, it turns out that the physical Hamiltonian is a \emph{non-linear} function of gravity and matter variables in almost all time gauges, and \emph{cannot} be written as,\footnote{The full Hamiltonian constraint is generally quadratic in momenta, and hence any time variable made out of fields, would lead to a square root Hamiltonian.}
\be
\label{hamsplit}
\widehat{H}_p = \widehat{H}_G + \widehat{H}_M.
\ee
This means that it is not possible to identify what is a matter Hamiltonian, and since we dont even know what it is, it is not possible to compute its spectrum to get a matter-vacuum.

There is, however,  a very special choice of time in which it is possible to write down the above equation. This is the dust time gauge. (We will return to the details of this choice of time in sections (\ref{dtg}) and (\ref{dtgauge})) We assume that we are in this gauge. We can then define the matter-vacuum as in (\ref{matvacdyn}), and note that it is a state that lies in the full Hilbert space $\mathds{H}_{QG}$. 

Now that we have defined what we mean by `vacuum' (summarized in Table \ref{table1_whatvac}), we proceed to formulate and answer the question `does vacuum gravitate?' in the next section.

\begin{table}[h!]
\centering
\begin{tabular}{ | p{4cm} | p{6.5cm}| p{6.5cm} | } 
 \hline
 \multicolumn{3}{|c|}{\textbf{What is a vacuum?}} \\
 \hline
 \underline{Type of vacuum} & \underline{Wheeler-deWitt approach}\nl & \underline{Physical Hamiltonian approach}\nl \\
 \hline
 Full vacuum\nl $\ket{0_{QG}} \in \mathds{H}_{QG}$ & Ambiguous since all states are annihilated by the Hamiltonian constraint operator. & Can be defined as,\nl $\widehat{H}_p(\hat{g},\hat{\Phi}) \ket{0_{QG}} = E_0 \ket{0_{QG}}$. \\
 \hline
 Matter vacuum on a dynamical background ($\hat{g}$)\nl $\ket{0_{M}} \in \mathds{H}_{QG}$ & Not known. Has to be a simultaneous eigenstate of both $\widehat{H}_G(\hat{g})$ and $\widehat{H}_M(\hat{g},\hat{\Phi})$. & Not possible in general. Can only be done in the dust-time gauge,\nl $\widehat{H}_M^{DT}(\hat{g},\hat{\Phi}) \ket{0_{M}} = E_0 \ket{0_{M}}$. \\ 
 \hline
 Matter vacuum on a fixed background ($g$)\nl $\ket{\widetilde{0}_{M}} \in \mathds{H}_{M}$ & A matter vacuum similar to the QFT vacuum can be defined using the matter Hamiltonian. However, the Wheeler-deWitt equation still has to be solved. This is similar to the semiclassical EFEs. & Not possible in general. Can only be done in the dust-time gauge,\nl $\widehat{H}_M^{DT}(g,\hat{\Phi}) \ket{\widetilde{0}_{M}} = \widetilde{E}_0 \ket{\widetilde{0}_{M}}$. \\ 
 \hline
\end{tabular}
\caption{A summary of section \ref{what_vac}: `What is a vacuum?'}
\label{table1_whatvac}
\end{table}

\begin{table}[h!]
\centering
\begin{tabular}{ | p{4cm} | p{6.5cm}| p{6.5cm} | } 
 \hline
 \multicolumn{3}{|c|}{\textbf{Does vacuum gravitate?}} \\
 \hline
 \underline{Type of vacuum} & \underline{Wheeler-deWitt approach}\nl & \underline{Physical Hamiltonian approach}\nl \\
 \hline
 Full vacuum\nl $\ket{0_{QG}} \in \mathds{H}_{QG}$ & No notion of a unique vacuum state, therefore this question cannot be addressed here. & Can calculate $ \bra{0_{QG}} \widehat{R}(\hat{g},\hat{\pi}) \ket{0_{QG}} $, however, the notion of gravitate is not clear in this context since $\ket{0_{QG}}$ is an eigenstate of the full Hamiltonian $\widehat{H}_p(\hat{g},\hat{\Phi})$, and lies in the full Hilbert space $\mathds{H}_{QG}$. \\
 \hline
 Matter vacuum on a dynamical background ($\hat{g}$)\nl $\ket{0_{M}} \in \mathds{H}_{QG}$ & Not known. If it exists, then one can calculate $ \bra{0_M} \widehat{R}(\hat{g},\hat{\pi}) \ket{0_M} $. The notion of gravitate is not clear here, since $\ket{0_M}$ is also an eigenstate of the gravitational Hamiltonian. & Only in the dust-time gauge, one can calculate $ \bra{0_M} \widehat{R}(\hat{g},\hat{\pi}) \ket{0_M} $. Not possible otherwise. \\ 
 \hline
 Matter vacuum on a fixed background ($g$)\nl $\ket{\widetilde{0}_{M}} \in \mathds{H}_{M}$ & Gravitate is not well-defined here, since a background has already been fixed, and therefore the curvature is also fixed. & Gravitate is not well-defined here, since a background has already been fixed, and therefore the curvature is also fixed. \\
 \hline
\end{tabular}
\caption{A summary of section \ref{vac_grav}: `Does vacuum gravitate?'}
\label{table2_vacgrav}
\end{table}

\section{Does vacuum gravitate?} \label{vac_grav}

A question that is often asked is `does vacuum gravitate?' This question is understood to mean that if we take some quantum matter which is in its vacuum state, how will it affect the geometry. Before answering this question, it is useful to understand the meaning of vacuum (more specifically, what it means to say `quantum fields in their vacuum state'). It is important to note here that in QFT (which does not include gravity), the vacuum state is background dependent.\footnote{In fact, it is coordinate dependent. Consider for instance the vastly different vacuum state obtained by doing a coordinate tranformation in flat space from Minkowski to Rindler coordinates.} As we saw in the previous section, it is non-trivial to define a vacuum within the full gravity-matter theory. Moreover, this question arises from the \emph{assumption} that there is a connection between flat/fixed background quantum field theory and general relativity, given by the semiclassical EFEs,
\be
G_{\mu \nu} + \Lambda g_{\mu \nu} = 8 \pi \bra{0_M} \widehat{T}_{\mu \nu} \ket{0_M},
\ee
where $\widehat{T}_{\mu \nu}$ is the quantized matter stress-energy tensor, and $\ket{0_M}$ is the `matter-vacuum'. One big problem with this approach (among others, see \S 2.2A of \cite{Isham:1995wr}, or \cite{PhysRevLett.47.979} for instance), is that this matter-vacuum state is defined on some fixed background metric (usually Minkowski). The very notion of a matter Hilbert space (and inner products) \emph{requires} the existence of some fixed metric. But in the EFEs, the spacetime metric is what we are solving for. A flat background is thus a \emph{requirement} to define the very equations we are solving for the background itself. This means that the semiclassical EFEs alone are not a consistent set of equations for dealing non-perturbatively with the full gravity-matter system.

It has been suggested that a final solution may only lie in a full non-perturbative theory of quantum gravity, which as of now, we don't know. However, an argument can be made just by considering some general assumptions about a non-perturbative theory of quantum gravity.

Following from the previous section, there is only one case where the vacuum of the full gravity-matter theory can be defined, which is the physical Hamiltonian approach. Let us call this vacuum $\ket{0_{QG}}$. Now to answer the question whether or not it `gravitates', we need to understand what is meant by that term. The term `gravitate' is usually understood as `to curve spacetime' i.e. given this vacuum state, how does it affect spacetime (how does it affect the gravitational variables). 

In this non-perturbative approach however, $\ket{0_{QG}}$ is the vacuum of the full gravity and matter sector, and lives in the full Hilbert space $\mathds{H}_{QG}$. It is calculated \emph{using} the dynamical gravitational variables, and hence the term gravitate is meaningless here. One can presumably define the 4-Ricci scalar operator $\widehat{R}(\hat{g},\hat{\pi})$ and calculate its expectation value in this state: $ \bra{0_{QG}} \widehat{R}(\hat{g},\hat{\pi}) \ket{0_{QG}} $, which would give some sort of an average geometrical curvature in this state, but whatever it turns out to be, zero or non-zero, it is fully consistent in this non-perturbative theory. There is no such thing as a `matter backreaction' here, this approach already includes all `backreactions' since it is non-perturbative, and gravity is dynamical. Now let us try to answer the question whether the matter-vacuum gravitates.

\subsection{Does matter-vacuum gravitate?}

From the previous section, we know that it is not possible in general to define a matter-vacuum in either the Wheeler-deWitt approach,\footnote{Suppose, for the sake of argument that a matter-vacuum state as in (\ref{matgravvacwdw}) can be defined. The meaning of the question `does matter-vacuum gravitate', is not clear here since the matter vacuum is also simultaneously an eigenstate of the gravity Hamiltonian. In general, this matter-vacuum state will depend on the dynamical gravitational degrees of freedom, and hence is distinct from the matter-vacuum state of QFT on a fixed background. One could still calculate an expectation value of the 4-Ricci scalar in this state, which would provide (fully consistently, without worrying about any `back-reactions') a measure of spacetime curvature in this state.} or the physical Hamiltonian approach. Hence, the question of backreaction of a matter-vacuum is not well-posed, since a matter-vacuum cannot even be identified. 
There is, however, one exception within the physical Hamiltonian approach in which a matter-vacuum can be defined: the dust time gauge. For the purpose of this section then, we assume that we are in this gauge. The results presented below will hold \emph{only} in this time gauge.\footnote{Or any other time gauge which allows (\ref{hamsplit}) to hold. To date, dust time gauge is the only one known which does this.} The full physical Hamiltonian becomes: $ \widehat{H}_p = \widehat{H}_G(\hat{g}) + \widehat{H}_M(\hat{g},\hat{\Phi})$. There are two possibilities [(\ref{matvacfix}) and (\ref{matvacdyn})] for the matter-vacuum state:

\begin{itemize}

\item Fix a background, which means that the gravitational degrees of freedom ($g$) are no longer dynamical. This means that a matter-vacuum state lying purely in the matter Hilbert space (although still dependent on the fixed background) can be defined. But once we have fixed the background, the curvature is also fixed. There is no backreaction, by construction. Therefore the question `Does matter-vacuum curve spacetime?' is meaningless here. The notion that we calculate vacuum energy density on some fixed background and then find that it is huge and hence should curve spacetime immensely is \emph{inconsistent} here since we have already fixed a background and no backreaction is allowed. To include backreaction, we have to go the full theory where gravity is dynamical.

\item Instead of fixing the background, find the lowest energy eigenstate of the matter Hamiltonian $\widehat{H}_M(\hat{g},\hat{\Phi})$ (which would lie in the full Hilbert space) and call that the matter-vacuum, i.e $\ket{0_M} \in \mathds{H}_{QG}$. It is important here to note the distinction, that this is a ``matter'' vacuum state which lies in the full Hilbert space and also depends on the \emph{dynamical} gravitational degrees of freedom $\hat{g}$. This vacuum is quite different from the QFT vacuum on a fixed background. Furthermore, this matter-vacuum state will be, in general, distinct from the vacuum state $\ket{0_{QG}}$ of the full theory which satisfies,
\be
\widehat{H}_p \ket{0_{QG}} = \Big( \widehat{H}_G(\hat{g}) + \widehat{H}_M(\hat{g},\hat{\Phi}) \Big) \ket{0_{QG}} = E_0 \ket{0_{QG}},
\ee
therefore we can call it a matter-vacuum.

Now that we have this state, we can compute the curvature of spacetime by taking an average of the 4-Ricci scalar in this state: $ \bra{0_M} \widehat{R}(\hat{g},\hat{\pi}) \ket{0_M} $ (Or the Riemann tensor in general, or any other natural object to measure curvature). This calculation would then be an answer to the question whether the matter-vacuum gravitates. Note that as mentioned earlier, this calculation can only be done in the dust time gauge.

\end{itemize}

(A summary of this section can be found in Table \ref{table2_vacgrav}.) Our approach here will be to choose a time function on the classical phase space, reduce to a physical Hamiltonian (for the full non-perturbative matter-gravity) system and then define the vacuum as the ground state of this full system (\emph{and not just a matter-vacuum}). \emph{This is a fully non-perturbative formalism, in which both gravity and matter are fully dynamical.} We will \emph{not} fix a background, but will keep it dynamical (as \emph{opposed} to QFT on a fixed background). We will then look at the relation between the vacuum energy density and the cosmological constant. In certain time gauges where it is difficult to quantize, we will leave the result at a classical level. This will not change our argument since the form of $\rho_{vac}$ as a function of $\Lambda$ will carry over to the quantized case, and the fact will still remain true that the result is to be viewed as a \emph{calculation} for $\rho_{vac}$ for a \emph{given} value of $\Lambda$.

\section{Vacuum energy density and the cosmological constant} \label{classical_vac}

We now look at the relation between energy density and the cosmological constant in three different time gauges.

\subsection{Scalar field time}

In the presence of a scalar field, we can construct a clock made out of matter degrees of freedom. One way to do that is to consider the scalar field itself as time \cite{Rovelli:1993bm},
\be
\label{phitime}
t = \phi.
\ee
The $\p_a \phi$ terms vanish from the Hamiltonian and the diffeomorphism constraint ($\partial_a \phi = \partial_a t = 0$). The time gauge fixing condition has to be dynamically preserved, which means that,
\be
\label{sf_time_fix}
\dot{t} = \dot{\phi} = \{t, H\} = 1,
\ee
where,
\be
H = \int d^3x \Big[ N \Big(\H_G + \H_{\phi}\Big) + N^a \Big(C_a^G + C_a^{\phi} \Big) \Big]
\ee
is the full Hamiltonian constraint with the scalar and the diffeo part given by (\ref{phi_ham}) and (\ref{start_ham}) respectively. Calculating this Poisson bracket then fixes the lapse function,
\bea
\{t,H\} &=& \{\phi,H\} = \dfrac{\delta H}{\delta \widetilde{\PP}} \nn \\
        &=& N \dfrac{\widetilde{\PP}}{\sq} + N^a \p_a \phi = N \dfrac{\widetilde{\PP}}{\sq} = 1 \nn \\
				&\Rightarrow& N = \dfrac{\sq}{\widetilde{\PP}},
\eea
where in the second step we have used $\p_a \phi = \p_a t = 0$ and (\ref{sf_time_fix}).
The variable conjugate to our choice of time is $\widetilde{\PP}$, since $\{\phi,\widetilde{\PP}\} = 1$, which makes our physical Hamiltonian: $\H_p = -\widetilde{\PP}$. Solving the Hamiltonian constraint (\ref{phi_ham}) for $\widetilde{\PP}$ then gives,
\be
\widetilde{\PP} = \pm \sqrt{\pi^2 - 2 \pi^{ab} \pi_{ab} + 2 q (R - \Lambda) - q m^2 t^2}.
\ee
We choose the appropriate sign here (and in future) to keep the physical Hamiltonian (density) positive,
\be
\H_p = \sqrt{\pi^2 - 2 \pi^{ab} \pi_{ab} + 2 q (R - \Lambda) - q m^2 t^2}.
\ee

We can find the (time) gauge fixed action by using $\H_p = -\widetilde{\PP}$, $\H_G + \H_\phi = 0$, $C_a^{\phi} = 0$ and (\ref{sf_time_fix}) in the full action (\ref{start_action})
\be
S^{GF} = \int d^3x dt \Big[ \pi^{ab} \dot{q}_{ab} - \H_p - N^a C_a^G \Big].
\ee

Note that to keep the Hamiltonian (density) real, we must have that,
\be
t^2 \leq \dfrac{1}{m^2} \Bigg[ \dfrac{\pi^2(x,t) - 2\pi^{ab}(x,t) \pi_{ab}(x,t)}{q(x,t)} + 2 (R(x,t)-\Lambda) \Bigg].
\ee
This equation is to be viewed as an \emph{implicitly} defined upper bound for time, beyond which the scalar field time gauge breaks down. Note that it also depends on the spatial point, $x$. (In the massless limit, $m \rightarrow 0$, this gauge is fine since the domain of $t$ becomes the whole real line.) Along with this, since we also want to keep time real, we must have that,
\be
\dfrac{\frac{1}{2}\pi^2 - \pi^{ab} \pi_{ab}}{q} + R \geq \Lambda.
\ee
These are just the conditions in which our time gauge is valid, and tell that, in general, the scalar field time gauge is not a particularly good choice.\footnote{See \cite{PhysRevD.92.064031} for a recent study regarding the viability of scalar field as time. Compared to our case, they consider a massless scalar field and without a cosmological constant.}

For a Hamiltonian system with the physical Hamiltonian $H_p = \int d^3x ~ \H_p$, in a volume $V$, the energy density is naturally \emph{defined} as,
\be
\rho \equiv \dfrac{1}{V} \int d^3x ~ \H_p.
\ee
For our case, since we have the physical Hamiltonian for the full gravity-matter system, $\rho$ is the energy density for the fully coupled non-perturbative gravity-matter system. It is \emph{not} just the matter energy density and is \emph{not} to be confused with other energy densities, e.g. $\rho$ appearing in the stress energy tensor of a perfect fluid or $\rho$ appearing in the Friedmann equations. In general (as we will see below), it is also \emph{not} the energy density of matter plus the energy density of gravity. It is the energy density of the full inextricably mixed gravity-matter system, which comes from a physical Hamiltonian after identifying a time function.
 
Returning back to scalar field time, the energy density is calculated as,
\bea
\rho &=& \dfrac{1}{V} \int d^3x ~\H_p \nn \\
     &=& \dfrac{\int d^3x \sqrt{\pi^2 - 2 \pi^{ab} \pi_{ab} + 2 q (R - \Lambda) - q m^2 t^2}}{\int d^3x \sq}.
\eea
This shows that the energy density is not linearly proportional to the cosmological constant, rather $\rho \propto \sqrt{\Lambda}$. It also shows that $\rho$ is explicitly dependent on time and cannot be written as $\rho_{\mbox{matter}} + \rho_{\mbox{gravity}}$. We will see that these generic features will be true of almost all time gauges.

\subsection{York time}

In this section, we switch off the scalar field for simplicity. A time variable constructed entirely out of gravitational degrees of freedom is the York time gauge,\footnote{This time gauge plays a central role in shape dynamics. See \cite{Mercati:2014ama,Barbour:2013goa}.}\cite{PhysRevLett.28.1082}
\be
\label{ytime}
t = \dfrac{2}{3} \dfrac{\pi}{\sq} = \dfrac{2}{3} \dfrac{q_{ab} \pi^{ab}}{\sq}.
\ee
The momentum conjugate to this time variable is, $p_t = \sq$ since $ \{t,p_t\} = \bigg\{\dfrac{2}{3}\dfrac{\pi}{\sq} , \sq \bigg\} = 1. $ Therefore, the physical Hamiltonian (density) is,
\be
\H_p = \sq.
\ee
For our purpose here, to calculate the energy density $\rho$ (as a function of $\Lambda$), we don't need to solve the Hamiltonian constraint for this physical Hamiltonian in terms of the reduced degrees of freedom since,\footnote{A technique to solve the full Hamiltonian constraint and the diffeomorphism constraint in York time is to consider conformal transformations of the metric: $q_{ab} \rightarrow \phi^4 q_{ab}$ (here, $\phi$ is not the scalar field, but the conformal factor) under which the Hamiltonian constraint becomes the Lichnerowicz-York(LY) equation \cite{Barbour:2013goa}:
\be
8 \dfrac{\nabla^2 \phi}{\phi^5} + \dfrac{\pi_{ab}^T \pi^{ab}_T}{\phi^{12} q} - \dfrac{R}{\phi^4} - \dfrac{3}{8}t^2 + \Lambda = 0. \nn
\ee
Here $ \pi_{ab}^T \equiv \pi_{ab} - \frac{1}{3} \pi q_{ab} $ is the traceless part of the gravitational momentum. Once this equation is solved for $\phi$, the physical Hamiltonian is given as,
\be
H = \int d^3x \sq \phi^6. \nn
\ee}
\be
\label{yrho}
\rho =  \dfrac{1}{V} \int d^3x ~\H_p = \dfrac{\int d^3x \sq}{\int d^3x \sq} = 1.
\ee
Restoring units we get,\footnote{Since we are in the classical theory, the only parameters we have are $G,c$ and $\Lambda$. All dimensions have to be constructed out of these. For York time (\ref{ytime}) to be dimensionally correct ($\pi$ has units 1/length and $q$ is dimensionless), we must have that $t =\frac{1}{c \Lambda} \frac{2}{3} \frac{\pi}{\sq}$.} 
\be
\label{yorkrho}
\rho = \dfrac{\Lambda c^2}{8 \pi G}.
\ee
Interestingly, the energy density is independent of (almost) everything! It is clear from (\ref{yrho}) that even if we add any matter in the system, the energy density would still remain constant. It is time independent and is linearly proportional to $\Lambda$ due to our choice of units. Note that the vacuum energy density calculated above is exactly the same as the one calculated from EFEs (\ref{EFE2}).

\subsection{Dust time} \label{dtg}

Finally, we consider another matter clock made from a dust field \cite{PhysRevLett.108.141301}. The dust Hamiltonian is
\be
\H_D = \sqrt{\widetilde{P}_T^2 + q^{ab} C_a^D C_b^D },
\ee
where,
\be
C_a^D = -\widetilde{P}_T \partial_a T,
\ee
and $(T, \widetilde{P}_T)$ are the dust field phase space variables. This is to be added to eq (\ref{start_action}) to obtain the full theory. The corresponding symplectic term is $\widetilde{P}_T \dot{T}$.
Our choice of time is the level surfaces of dust,
\be
t = T.
\ee
The dust diffeomorphism constraint vanishes ($\partial_a T = \partial_a t = 0$) and the physical Hamiltonian (density) becomes,
\bea
\H_p &=& \dfrac{1}{\sq} \Big(\pi^{ab} \pi_{ab} - \frac{1}{2} \pi^2 \Big) + \sq ( \Lambda - R) \nn \\
    &+& \dfrac{\widetilde{\PP}^2}{2 \sq} + \dfrac{1}{2} \sq q^{ab} \partial_a \phi \partial_b \phi + \dfrac{1}{2} \sq m^2 \phi^2.
\eea
Notice that quite remarkably, this physical Hamiltonian (density) is just the old Hamiltonian constraint without the dust field. In this sense, dust-time seems to be a good time gauge. The physical Hamiltonian (density) is not a square root and is also not explicitly time dependent.

The gauge fixed action is,
\be
S^{GF} = \int d^3x dt \Big[ \pi^{ab} \dot{q}_{ab} + \widetilde{\PP} \dot{\phi} - \H_p - N^a(D_b \pi^b_a + \PP \partial_a \phi) \Big].
\ee
The energy density becomes,
\bea
\rho &=& \dfrac{1}{\int d^3x \sq} \int d^3x \Big[ \dfrac{1}{\sq} \Big(\pi^{ab} \pi_{ab} - \frac{1}{2} \pi^2 \Big) + \sq ( \Lambda - R) \nn \\
    &+& \dfrac{\widetilde{\PP}^2}{2 \sq} + \dfrac{1}{2} \sq q^{ab} \partial_a \phi \partial_b \phi + \dfrac{1}{2} \sq m^2 \phi^2 \Big].
\eea
It is clear from here that $\rho$ is linear in $\Lambda$ and is also not explicitly time dependent. Again, this equation is to be viewed as a calculation for the energy density given a value of $\Lambda$.

\section{Reduction to Cosmology} \label{cosmo}

The full theory analysis in the previous section was done classically and we did not quantize due to difficulties with quantizing full general relativity. In this section, we reduce from full GR to a homogeneous, isotropic cosmology. This reduction to cosmology provides a concrete instance where we can actually go ahead and perform a full quantization\footnote{For a pioneering study on quantizing the Friedmann universe with different choices of time, see \cite{PhysRevD.11.768}.} (in some time gauges) and explicitly calculate the vacuum energy density in the full (homogeneous) quantum theory. We will see that the general features of the full theory classical analysis still hold here in the (symmetry-reduced) quantum theory: the vacuum energy density is, in general, time dependent and a non-linear function of $\Lambda$.

In addition to what was done in \cite{PhysRevLett.116.061302}, here we also include a non-zero spatial curvature of the universe, perform a Fermionic quantization of the physical Hamiltonian for their choice of time gauge, and calculate the vacuum energy density for other choices of time as well. We see that introducing a positive spatial curvature restricts the domain of the time function, Fermionic quantization produces (almost) the same result as usual quantization, and there is a time choice (dust time), in which the energy density \emph{is} a linear function of $\Lambda$, in \emph{contrast} to the results of \cite{PhysRevLett.116.061302}.

We now reduce to a Friedmann-Robertson-Walker (FRW) universe with the ansatz
\be
q_{ab} = a^2(t) h_{ab} , ~ \pi^{ab} = \dfrac{P_a(t)}{6a(t)} h^{ab} \sh,
\ee
where $a(t)$ is the scale factor, $P_a$ is the momentum canonically conjugate to the scale factor, $h_{ab} = \dfrac{1}{f^2(r)} e_{ab}$ is an auxiliary metric (which provides the density weight), $e_{ab} = \text{diag}(1,1,1)$ is the flat metric,
\be
f(r) = 1 + \dfrac{\ka r^2}{4},
\ee
and $\ka$ is the spatial curvature and can be either 0 (flat), +1 (closed) or -1 (open). 
(This is the isotropic form of the FRW line element. See \S 149 in \cite{tolman1987relativity}.)

We also assume that the scalar field is homogeneous,\footnote{For the dust field, assuming homogeneity i.e $T = T(t), ~ \widetilde{P}_T = \sh ~ P_T(t)$ means that the dust diffeomorphism contraint vanishes ($C_a^D = 0$) and the Hamiltonian reduces to: $ H_D = \widetilde{P}_T$.}
\be
\phi = \phi(t) ~ , ~ \widetilde{\PP} = \sh ~ \PP(t).
\ee

With this homogeneity ansatz, the diffeomorphism constraint vanishes identically and we are left with, (after performing the spatial integration)
\be
S = V_p \int dt ~ \Bigg( P_a \dot{a} + \PP \dot{\phi} - NH \Bigg),
\ee
with the Hamiltonian constraint,
\be
\label{cosmo_ham}
H =  \dfrac{-P_a^2}{24 a}  + a^3 \Bigg( \Lambda - \dfrac{6 \ka}{a^2} \Bigg)  + \dfrac{\PP^2}{2 a^3} + \dfrac{1}{2} a^3 m^2 \phi^2,
\ee
where $ V_p = \displaystyle \int \sh ~ d^3x $. We will now fix different time gauges and calculate the vacuum energy density.

\subsection{The volume time gauge} \label{voltime}

First we choose the time gauge considered in \cite{PhysRevLett.116.061302}. Here we generalize their result to arbitrary spatial curvature and also perform a Fermionic quantization of the resultant Hamiltonian.

Time is chosen to be the (spatial) volume of the universe, a very natural choice for us since we are currently living in an expanding universe and hence the volume constitutes a monotonically increasing function.
\be
t = \int d^3x \sq = V_p a^3.
\ee
With scale invariant variables defined as: $\pp = V_p \PP$ and $ \kb = V_p^{\frac{2}{3}} \ka $ we get the gauge fixed action,
\be
S^{GF} = \int dt \Big[ \pp \dot{\phi} - H_p \Big],
\ee
\be
\label{voltimeham}
H_p = \sqrt{ \dfrac{8}{3} \Bigg[ \Lambda - \dfrac{6 \kb}{t^{2/3}} + \dfrac{\pp^2}{2t^2} + \dfrac{1}{2} m^2 \phi^2 \Bigg] }.
\ee
Note that to keep the Hamiltonian hermitian, we must have that
\be
\label{Ham_hermitian}
\Lambda - \dfrac{6 \kb}{t^{2/3}} + \dfrac{\pp^2(t)}{2t^2} + \dfrac{1}{2} m^2 \phi^2(t) \geq 0,
\ee
given that $\Lambda \geq 0$, the above equation is automatically satisfied for $\kb = 0$ and $\kb < 0$, however, for $\kb >0$, the above equation is to be viewed as a restriction on time $t$, beyond which our time gauge breaks down (Note that since $\pp$ and $\phi$ are functions of time, this is an implicitly defined equation for time.)

The energy density can be calculated as,
\be
\rho = \dfrac{H_p}{V_{physical}} = \dfrac{H_p}{\int d^3x \sq} = \dfrac{H_p}{V_p a^3} = \dfrac{H_p}{t}.
\ee
Now that we have the classical Hamiltonian (and energy density), we consider two different ways of quantization: usual (Schrodinger) quantization; and Fermionic quantization (a'la Dirac).

\subsubsection{Standard quantization}

The Hamiltonian (\ref{voltimeham}) is the square root of a shifted harmonic oscillator. Following \cite{PhysRevLett.116.061302}, this can be readily quantized. For any operator $\hat{A}$ that has a \emph{positive} spectrum $a_n$, the spectrum of the square root of that operator $\sqrt{\hat{A}}$ is given by $\sqrt{a_n}$. To compute the spectrum for our Hamiltonian, we require that the argument inside the square root be positive. But that is already required to keep the Hamiltonian hermitian (and real) from (\ref{Ham_hermitian}). The term inside the square root is,
\be
\dfrac{8}{3} \Bigg[ \Lambda - \dfrac{6 \kb}{t^{2/3}} + \dfrac{\pp^2}{2t^2} + \dfrac{1}{2} m^2 \phi^2 \Bigg]
\ee 
which is just a time-dependent, shifted harmonic oscillator. For our purpose here, since we are only interested in the spectrum and not in solving the time-dependent Schrodinger equation, we can treat $t$ as some parameter (like mass) in the Hamiltonian. Therefore, the shift above is a constant, and hence if we find the spectrum of the remaining part, the constant shift can just be added in the spectrum since, given an operator $\hat{B}$ with the eigenvalues $E_B$, the eigenvalues of another operator $\hat{B} + C\hat{I}$ (where $C$ is a constant and $\hat{I}$ is the identity operator), are just $E_B + C$. The remaining part looks like a harmonic oscillator. Recall that for a simple harmonic oscillator like,
\be
H_{SHO} = \dfrac{P^2}{2 \bar{m}} + \dfrac{1}{2} \bar{m} \omega^2 x^2,
\ee
the energy spectrum is given as,
\be
E_{SHO} = \Big(n + \dfrac{1}{2}\Big) \omega. 
\ee
where $n=0,1,2,...$ Comparing with our Hamiltonian here, if we set $\bar{m} = t^2$ and $\omega = m/t$, we get the spectrum as,
\be
E = \Big(n + \dfrac{1}{2}\Big) \dfrac{m}{t}. 
\ee
We can now combine all of the above results to get the spectrum for the full square root Hamiltonian (\ref{voltimeham}),
\be
E_n = \sqrt{ \dfrac{8}{3} \Bigg[ \Lambda - \dfrac{6 \kb}{t^{2/3}} + \Bigg( n + \dfrac{1}{2} \Bigg) \dfrac{m}{t}  \Bigg]}.
\ee
Finally, we can calculate the energy density, which is just (recalling that volume $=t$),
\be
\rho_n = \dfrac{E_n}{t} = \dfrac{1}{t} \sqrt{ \dfrac{8}{3} \Bigg[ \Lambda - \dfrac{6 \kb}{t^{2/3}} + \Bigg( n + \dfrac{1}{2} \Bigg) \dfrac{m}{t}  \Bigg] }.
\ee
Naturally for the vacuum, we choose $n=0$, which gives,
\be
\label{rho_voltime}
\rho = \dfrac{1}{t} \sqrt{ \dfrac{8}{3} \Bigg[ \Lambda - \dfrac{6 \kb}{t^{2/3}} + \dfrac{m}{2t}  \Bigg] }.
\ee
This equation is to be viewed as a \emph{calculation} for the value of the vacuum energy density (of the full gravity-matter system) given the observed value of the cosmological constant, hence making it clear that in this framework, there is no cosmological constant problem. We also note that the vacuum energy density is not linearly proportional to $\Lambda$ and is explicitly time dependent. 

\subsubsection{Fermionic quantization}

We can also perform a Fermionic quantization a'la Dirac. Consider a massless scalar field with the Hamiltonian
\be
H = \sqrt{ \dfrac{8}{3} \Bigg( \Lambda - \dfrac{6 \kb}{t^{2/3}} + \dfrac{\pp^2}{2t^2} \Bigg) },
\ee
and write it as linear in the momentum,\footnote{This trick is not possible for a massive scalar field since the mass term ($m^2 \phi^2$) and the momentum term ($\pp^2$) do not commute.}
\be
H = \alpha A + \beta \dfrac{\pp}{t} B,
\ee
where $\alpha = \sqrt{\dfrac{8}{3} \Bigg( \Lambda - \dfrac{6 \kb}{t^{2/3}} \Bigg)}$ , $\beta = \sqrt{\dfrac{8}{6}}$ and A and B are some matrices.

Then the condition that the square of this linear Hamiltonian be equal to the square of the original Hamiltonian leads to A and B being Pauli matrices. We choose the following two:
\[ A = \sigma_x =  \left( \begin{array}{ccc}
0 & 1 \\
1 & 0  \end{array} \right), \] 
and
\[ B = \sigma_z =  \left( \begin{array}{ccc}
1 & 0 \\
0 & -1  \end{array} \right). \] 
The Hamiltonian becomes,
\[ H = \left( \begin{array}{ccc}
\dfrac{\beta}{t} \pp & \alpha \\
 \alpha & -\dfrac{\beta}{t} \pp  \end{array} \right). \] 
To keep the Hamiltonian hermitian, we must have from the above that $ \alpha^* = \alpha $ which leads to, 
\be
 \Lambda \geq \dfrac{6 \kb}{t^{2/3}}.
\ee
We can view the above equation as introducing a lower bound on time $t = t_c$, such that, if $ t_c = \Big(\frac{6 \kb}{\Lambda} \Big)^{\frac{3}{2}} $, then the above equation is automatically satisfied for all $t \geq t_c$.

Now, we want to find the (instantaneous) eigenstates of this Hamiltonian,\footnote{Here, we are only interested in calculating the vacuum energy density. For a detailed treatment of this `spinorial' Hamiltonian (in a different time gauge) see \cite{Kreienbuehl:2009ub}.}
\be
\label{evaleqn}
\widehat{H}(t) \ket{\psi_n(t)} = E_n(t) \ket{\psi_n(t)}.
\ee
The general wavefunction at any time can be written as: $ \ket{\Psi} = \sum_n c_n(t) \ket{\psi_n(t)} $, from which we can then calculate the expectation value of the Hamiltonian,
\be
\expect{\widehat{H}} = \dfrac{\bra{\Psi(t)} \widehat{H}(t) \ket{\Psi(t)}}{ \braket{\Psi(t)}{\Psi(t)}} = \dfrac{\sum_n |c_n(t)|^2 E_n(t)}{\sum_n |c_n(t)|^2}.
\ee
In the position representation, $ \pp = -i \pphi $, eqn (\ref{evaleqn}) becomes \newline

\begin{center} $ \begin{bmatrix} -\dfrac{i \beta}{t} \pphi & \alpha \\ \alpha & \dfrac{i \beta}{t} \pphi \end{bmatrix} \left( \begin{array}{c} \psi_1 \\ \psi_2 \end{array} \right) = E_n(t) \left( \begin{array}{c} \psi_1\\ \psi_2 \end{array} \right), $ \end{center}

with the solutions,
\be
\psi_1 = e^{i k \phi} \Big( -C_1 + C_2 \Big) + e^{-i k \phi} \Big( C_1 + C_2 \Big),
\ee
and
\bea
\psi_2 &=& e^{i k \phi} \Bigg[ \Bigg(\frac{-C_1 + C_2}{\sqrt{8 \Big( \Lambda - \frac{6 \kb}{t^{2/3}} \Big)}} \Bigg) \Bigg( -\frac{2k}{t} + \sqrt{3} E \Bigg) \Bigg] \nn \\
&+& e^{-i k \phi} \Bigg[ \Bigg(\frac{C_1 + C_2}{\sqrt{8 \Big( \Lambda - \frac{6 \kb}{t^{2/3}} \Big)}} \Bigg) \Bigg( \frac{2k}{t} + \sqrt{3} E \Bigg) \Bigg],
\eea
where $C_1$ and $C_2$ are integration constants.

The eigenvalues are,
\be
E(t) = \pm \sqrt{\dfrac{8}{3} \Bigg( \Lambda - \dfrac{6 \kb}{t^{2/3}} + \dfrac{k^2}{2t^2} \Bigg)}.
\ee
Hence, we get,
\be
\expect{\widehat{H}} = \pm \sqrt{\dfrac{8}{3} \Bigg( \Lambda - \dfrac{6 \kb}{t^{2/3}} + \dfrac{k^2}{2t^2} \Bigg)},
\ee
and the energy density is
\be
\expect{\widehat{\rho}} = \pm \dfrac{1}{t} \sqrt{\dfrac{8}{3} \Bigg( \Lambda - \dfrac{6 \kb}{t^{2/3}} + \dfrac{k^2}{2t^2} \Bigg)}.
\ee
This turns out to be the same as before (Schrodinger quantization, with a massless scalar field). The only difference is that now, both positive and negative values are allowed. Once again, $\rho \propto \sqrt{\Lambda}$ and is explicitly time dependent.

We now consider three other time gauges within the homogeneous cosmological sector and see how the vacuum energy density changes.

\subsection{Scale factor time} \label{sf_time}

A geometrical time gauge related to volume time is the scale factor time,
\be
t = V_p^{\frac{1}{3}} a.
\ee
With the definitions $ \pp = V_p \PP $ and $ \kb = V_p^{\frac{2}{3}} \ka $, we get the gauge fixed action:
\be
S^{GF} = \int dt \Big( \pp \dot{\phi} - H_p \Big),
\ee
\be
H_p = t^2 \sqrt{ 24 \Big[ \Lambda - \dfrac{6 \kb}{t^2} + \dfrac{\pp^2}{2 t^6} + \dfrac{1}{2} m^2 \phi^2 \Big] }.
\ee
Again, for hermiticity, it is required that
\be
\Lambda - \dfrac{6 \kb}{t^2} + \dfrac{\pp^2}{2 t^6} + \dfrac{1}{2} m^2 \phi^2 \geq 0,
\ee
which is viewed as a restriction on $t$ beyond which our time gauge breaks down.

The energy density becomes
\be
\rho = \dfrac{H_p}{V_p a^3} = \dfrac{H_p}{t^3} = \dfrac{1}{t} \sqrt{ 24 \Big[ \Lambda - \dfrac{6 \kb}{t^2} + \dfrac{\pp^2}{2 t^6} + \dfrac{1}{2} m^2 \phi^2 \Big] },
\ee
again, this is a scaled and shifted harmonic oscillator and is readily quantized to give,
\be
\rho = \dfrac{1}{t} \sqrt{ 24 \Big[ \Lambda - \dfrac{6 \kb}{t^2} + \Big(n + \dfrac{1}{2}\Big) \dfrac{m}{t^3} \Big] }.
\ee
For vacuum, we choose $n=0$,
\be
\rho = \dfrac{1}{t} \sqrt{ 24 \Big[ \Lambda - \dfrac{6 \kb}{t^2} + \dfrac{m}{2t^3} \Big] }.
\ee
It is clear that this is different from what we had in volume time (\ref{rho_voltime}). Once again, we emphasize that the above equation is a \textit{calculation} for vacuum energy density given the cosmological constant. We also see that, as before, $\rho \propto \sqrt{\Lambda}$ and is explicitly time dependent.

\subsection{Scalar field time} \label{sfield_time}

Next, we use matter variables to make a clock and consider the scalar field itself as time,
\be
\label{phi_time}
t = \phi.
\ee
With scale invariant quantities defined as $ p_a = V_p^{\frac{2}{3}} P_a $ , $\ba = V_p^{\frac{1}{3}} a$ and $\kb = V_p^{\frac{2}{3}}\ka$, the gauge fixed action becomes,
\be
S^{GF} = \int dt \Big[ p_a \dot{\ba} - H_p \Big],
\ee
\be
\label{sfield_hamil}
H_p = \sqrt{ \dfrac{p_a^2 \ba^2}{12} - 2 \ba^6 \Big( \Lambda - \dfrac{6 \kb}{\ba^2} \Big) - \ba^6 m^2 t^2 }.
\ee
To keep the Hamiltonian hermitian, we must have,
\be
t^2 \leq \dfrac{1}{m^2} \Bigg( \dfrac{p_a^2}{12\ba^4} + \dfrac{12 \kb}{\ba^2} - 2 \Lambda \Bigg).
\ee
Also, since $t \in \mathds{R}$, we must have that,
\be
\Lambda \leq \dfrac{1}{\ba^4} \Bigg( \dfrac{p_a^2}{24} + 6 \kb \Bigg).
\ee
These conditions once again reflect the limited validity of the scalar field time gauge.
The energy density then becomes,
\be
\rho = \dfrac{H_p}{V_p a^3} = \dfrac{H_p}{\ba^3} = \sqrt{ \dfrac{p_a^2}{12 \ba^4} - 2 \Lambda + 12 \dfrac{\kb}{\ba^2} - m^2 t^2 }.
\ee
So far, this is classical. To quantize this theory, we look at our degrees of freedom, which are: $(a(t), p_a(t))$. So this is a 1-dimensional quantum mechanical system. We could impose the usual representations of these operators: $\hat{a} F(a) \equiv a F(a)$ and $\hat{p}_a F(a) \equiv -i \p_a F(a)$ where the functions $F(a)$ lie in some suitably constructed Hilbert space (to make sense of inner products, some sort of a measure has to be defined on the space of all $a$'s). Looking at the form of the physical Hamiltonian, there are two problems here: (i) To construct an operator out of $\frac{1}{\ba}$ and (ii) Operator ordering issues in making an operator out of the first term: $\dfrac{p_a^2}{12 \ba^4}$. Here, we will not try to solve these issues,\footnote{See \cite{Rovelli:1993bm} where the full theory (with a massless scalar field) is quantized in the scalar field time gauge.} but will state that at a classical level, we have $\rho \propto \sqrt{\Lambda}$ and is explicitly time dependent; generic features which are expected to be carried over to the quantum theory regardless of how the above problems are solved. This is analogous to what we have seen for our previous choices of time where we were able to quantize and these features remained valid in the quantum theory. 

\subsection{Dust time} \label{dtgauge}

Finally, we consider the dust time gauge,
\be
t = T.
\ee
We define, $ p_a = V_p^{\frac{2}{3}} P_a $, $\ba = V_p^{\frac{1}{3}} a$, $ \pp = V_p \PP $ and $\kb = V_p^{\frac{2}{3}}\ka$. The gauge fixed action is then,
\be
S^{GF} = \int dt \Bigg[ p_a \dot{\ba} + \pp \dot{\phi} - H_p \Bigg],
\ee
\be
\label{dtgauge_hamil}
H_p = \dfrac{-p_a^2}{24 \ba}  + \ba^3 \Bigg( \Lambda - \dfrac{6 \kb}{\ba^2} \Bigg)  + \dfrac{\pp^2}{2 \ba^3} + \dfrac{1}{2} \ba^3 m^2 \phi^2.
\ee

This Hamiltonian is not a square root Hamiltonian and is time-independent. There are no restrictions on the time parameter or the cosmological constant just to keep the Hamiltonian hermitian, implying that the dust field is a good choice of time. Comparing this with  (\ref{cosmo_ham}), we see that they are the same, the difference being that this is now a (non-vanishing) physical Hamiltonian, whereas (\ref{cosmo_ham}) was the Hamiltonian constraint. This is another feature of using the dust field as time.

For the quantum theory, we have the same problems here as discussed before in the case of scalar field time; defining inverse operators and operator ordering issues. Therefore, we will not proceed to quantize this theory, but will look at the energy density classically,
\be
\rho = \dfrac{H_p}{\ba^3} = \dfrac{-p_a^2}{24 \ba^4}  +  \Lambda - \dfrac{6 \kb}{\ba^2}  + \dfrac{\pp^2}{2 \ba^6} + \dfrac{1}{2} m^2 \phi^2.
\ee
Here, we see that the energy density is not explicitly time dependent and is linearly proportional to $\Lambda$. However, there is still no cosmological constant problem, since the above equation is to be viewed as a calculation for the energy density given a value of the cosmological constant.

\section{Inhomogeneities} \label{inhomo_cosmo}

At this stage, one may ask that the analysis in the previous section is for the homogeneous case, and is restricted to zero modes. How would it extend to inhomogeneities, and in particular, how does it relate to the understanding of the conventional cosmological constant problem (that `vacuum energy' is divergent due to infinite/large contributions coming from an infinite/large number of modes).

The purpose of this section is to look at the effect of inhomogeneities by performing a mode expansion around a spatially flat FRW background. We note that this is necessarily perturbative, since performing a mode expansion first requires the notion of a fixed background. This is because modes are defined as eigenfunctions of a Laplacian on a fixed background. In general, a mode expansion is neither possible (if the background is dynamical i.e $\hat{g}$), nor useful since the resulting momentum space Hamiltonian does not simplify, as we will see below.\footnote{See also \cite{Hollands:2004xv} for a discussion on problems with a mode treatment of QFT.}

For the first sub-section, we will sketch our arguments heuristically, since our purpose here is not to provide a detailed treatment of cosmological inhomogeneities, but to see the features of vacuum energy density. In the second sub-section, we repeat the calculation for the inhomogeneous model considered in \cite{PhysRevLett.116.061302}, but this time in a different time gauge to see explicitly that vacuum energy density (even in the inhomogeneous sector) does depend on the choice of time, as one would already have expected from the full theory analysis done in Sec (\ref{classical_vac}), and that there is no cosmological constant problem arising from a large number of modes.

\subsection{Cosmological perturbations}

Following \cite{0264-9381-11-2-011}, we expand the full Hamiltonian constraint (\ref{start_ham_full}) around a spatially flat FRW background, upto second order in perturbations. The degrees of freedom $ X =  (q_{ab}, \pi^{ab}, \phi, \PP)$ are split as,

\be
X(t,x) = X^{0}(t) + \delta X(t,x),
\ee

where $X^{0}(t)$ is the homogeneous background value, and $\delta X(t,x)$ is the perturbation. We will omit all $0$'s in the superscript in background variables for simplicity, with the understanding that any variables not appearing with a $\delta$ in front of them are homogeneous background variables.

The Hamiltonian constraint then reads,
\bea
\label{Ham_second_order}
\mathcal{H} &=& -\dfrac{P_a^2}{24a} + a^3 \Lambda + \dfrac{\PP^2}{2a^3} + \dfrac{1}{2} a^3 m^2 \phi^2 \nn \\
            &+& \Bigg( -\dfrac{P_a^2}{144a} + \dfrac{1}{2} a^3 \Lambda - \dfrac{\PP^2}{4a^3} + \dfrac{1}{4} a^3 m^2 \phi^2 \Bigg) q^{ab} \delta q_{ab} - a^3 \Bigg( q^{ac} q^{bd} - q^{ad} q^{bc} \Bigg) \partial_b \partial_c \delta q_{ad} - \dfrac{P_a}{6a^2}q_{ab} \delta \pi^{ab} + \dfrac{\PP}{a^3} \delta \PP + a^3 m^2 \phi \delta \phi \nn \\
						&+& \dfrac{1}{a^3} \Bigg( q_{ac} q_{bd} - \dfrac{1}{2} q_{ab} q_{cd} \Bigg) \delta \pi^{ab} \delta \pi^{cd} + \dfrac{(\delta \PP)^2}{2 a^3} + \dfrac{1}{2} a^3 m^2 (\delta \phi)^2  + \dfrac{1}{2} a^3 q^{ab} \partial_a \delta \phi \partial_b \delta \phi + f(\delta q_{ab}) \approx 0,
\eea
where $f(\delta q_{ab})$ is some second-order function of the metric perturbation, which does not have a $P_a$ in it (since it comes from the perturbation of the 3-Ricci scalar $R$), we will not need its exact form here. We \emph{define} our theory to be,
\be
S = \int d^3x dt [ P_a \dot{a} + \delta\pi^{ab} \dot{\delta q}_{ab} + \PP \dot{\phi} + \delta\PP \dot{\delta \phi} - N \mathcal{H} - N^a \mathcal{C}_a],
\ee
with $(a,P_a)$ and $(\phi,\PP)$ the homogeneous canonically conjugate variables, and $(\delta q_{ab},\delta \pi^{ab})$ and $(\delta \phi,\delta\PP)$ the inhomogeneous canonically conjugate variables. Here $\mathcal{C}_a$ is the diffeomorphism constraint expanded upto second order in perturbations whose exact form will not concern us here.

There are two ways to proceed from here, first is to fix a time gauge first, and then do a mode expansion, second is to first do a mode expansion, and then fix the time gauge. We look at both of these separately.

\subsubsection{Gauge fixing before mode expansion}

We fix the scale factor time gauge $t=a$ (ignoring factors of $V_0$ for simplicity). Then, solving the Hamiltonian constraint (\ref{Ham_second_order}) for its conjugate momentum $P_a$ gives the physical Hamiltonian density as,
\be
\mathcal{H}_p = \dfrac{ -\frac{12}{t} q_{ab} \delta \pi^{ab} \pm \sqrt{ (\frac{12}{t} q_{ab} \delta \pi^{ab})^2 + (6 + q^{ab}\delta q_{ab}) (144 t ~g(t,\phi,\PP,\delta q_{ab},\delta \pi^{ab},\delta \phi,\delta\PP)) } }{(6 + q^{ab}\delta q_{ab})},
\ee
where,
\bea
g(t,\phi,&\PP&,\delta q_{ab},\delta \pi^{ab},\delta \phi,\delta\PP) = t^3 \Lambda + \dfrac{\PP^2}{2t^3} + \dfrac{1}{2} t^3 m^2 \phi^2 \nn \\
            &+& \Bigg(\dfrac{1}{2} t^3 \Lambda - \dfrac{\PP^2}{4t^3} + \dfrac{1}{4} t^3 m^2 \phi^2 \Bigg) q^{ab} \delta q_{ab} - t^3 \Bigg( q^{ac} q^{bd} - q^{ad} q^{bc} \Bigg) \partial_b \partial_c \delta q_{ad} + \dfrac{\PP}{t^3} \delta \PP + t^3 m^2 \phi \delta \phi \nn \\
						&+& \dfrac{1}{t^3} \Bigg( q_{ac} q_{bd} - \dfrac{1}{2} q_{ab} q_{cd} \Bigg) \delta \pi^{ab} \delta \pi^{cd} + \dfrac{(\delta \PP)^2}{2 t^3} + \dfrac{1}{2} t^3 m^2 (\delta \phi)^2  + \dfrac{1}{2} t^3 q^{ab} \partial_a \delta \phi \partial_b \delta \phi + f(\delta q_{ab}),
\eea
and the physical Hamiltonian is given as,
\be
H_p = \int d^3x \mathcal{H}_p.
\ee

The inhomogeneous degrees of freedom $\delta X$ are expanded into modes as,
\be
\label{mode_expand}
\delta X(t,x) = \int \dfrac{d^3k}{(2\pi)^3} \delta X_{\kk}(t) e^{i \kk.\textbf{x}}.
\ee
It is clear from the square root form of the physical Hamiltonian, that the mode expansion will result in something like (heuristically),
\be
H_p = \int d^3x \sqrt{ \int \int \dfrac{d^3k}{(2\pi)^3} \dfrac{d^3k'}{(2\pi)^3} \delta X_{\kk}(t) \delta X_{\kk'}(t) e^{i (\kk+\kk').\textbf{x}} + ...  },
\ee
and since $\int d^3x$ is sitting outside the square root, it is not possible to carry out the $x$ integration to get the usual result. Hence a mode expansion is not very useful in this situation, since it is not clear how to proceed to a quantum theory from here. We will not consider this case further.

\subsubsection{Gauge fixing after mode expansion}

The other possibility is to carry out a mode expansion first, and then gauge fix. In this case, the Hamiltonian (before gauge-fixing) is,
\be
H = \int d^3x (N(t,x) \mathcal{H} + N^a(t,x) \mathcal{C}_a),
\ee
where $\mathcal{H}$ is given as in (\ref{Ham_second_order}). Since this is not a square root, the $x$ integration can be carried out. The usual way to proceed from here is to make the gauge choices that $N(t,x)=1$ and $N^a(t,x) = 0$ (e.g in \cite{0264-9381-11-2-011}). It is important to note that without these simplifying assumptions, the $x$ integration cannot be carried out here as well due to the arbitrary functions of $x$ sitting in the Hamiltonian (which would result in a convolution, instead of the usual delta function). Moreover, since we are interested in fixing a time-gauge later on, we have to check if the above choices for the lapse and shift are consistent. This is because fixing a time gauge restricts the allowed lapse functions (similarly, choosing a spatial gauge would fix the shift).

Looking ahead, we want to fix the scale factor gauge $t=a$, so we check first what possibilities of lapse are allowed. Since the gauge fixing condition has to be dynamically preserved, we must have that,
\be
\dot{t} = \{t,H\} = \Big\{a, \int d^3x (N(t,x) \mathcal{H} + N^a(t,x) \mathcal{C}_a)\Big\} = 1.
\ee

Calculating this Poisson bracket and using the explicit form of $\mathcal{H}$ from (\ref{Ham_second_order}), we get,
\be
\int d^3x \Bigg[ N(t,x) \Bigg( -\dfrac{P_a}{12a} - \dfrac{P_a}{72 a}q^{ab}\delta q_{ab} - \dfrac{1}{6a^2} q_{ab} \delta \pi^{ab} \Bigg) + N^a(t,x) \dfrac{\delta \mathcal{C}_a}{\delta P_a} \Bigg] = 1.
\ee

Since we used the homogeneous variable $a$ to fix the time gauge, the above equation is a global equation for $N(t,x)$ and $N^a(t,x)$. Also, since we are not fixing a spatial gauge here, we can set $N^a(t,x) = 0$ for simplicity (Although in the case of a full gauge fixing, one has to be careful whether this choice is consistent or not with the chosen gauge). Since our purpose is to be able to carry out the $x$ integration after the mode expansion, we require the lapse to be independent of $x$. Given that we have a global equation for $N(t,x)$, this is a consistent choice (although it is certainly not the full solution). Defining,
\be
F(t) = \int d^3x \Bigg( -\dfrac{P_a}{12a} - \dfrac{P_a}{72 a}q^{ab}\delta q_{ab} - \dfrac{1}{6a^2} q_{ab} \delta \pi^{ab} \Bigg),
\ee
we get the solution,
\be
\label{N_as_F}
N(t) = \dfrac{1}{F(t)},
\ee
given that $F(t)$ is nowhere zero. Note that this is not necessarily equal to one, and hence making the choice $N=1$ at the beginning would have been inconsistent for this scale factor time gauge.

We can now perform the mode expansion (\ref{mode_expand}) for all the inhomogeneous degrees of freedom and plug that into the full Hamiltonian constraint (\ref{Ham_second_order}). Again, we will not do the detailed calculation here, but the resultant Hamiltonian constraint after using (\ref{N_as_F}), $N^a(t,x)=0$, and carrying out the integration over $x$ will look something like,
\be
H_{\kk}(\Lambda; a,P_a,\phi,\PP,\delta q_{ab},\delta \pi^{ab}, \delta \phi, \delta \PP) = N(t) \Bigg[ V_0 (\mbox{homogeneous degrees}) + \int d^3k ~ ( \delta X_{\kk} + ...) \Bigg] \approx 0.
\ee

Once we fix the time gauge $t=a$, and solve the above for its conjugate momentum $P_a$ to get the physical Hamiltonian, it will be: (a) time dependent, since the Hamiltonian constraint above has $a$'s in it which will be replaced by $t$, (b) a square root, since $P_a$ appears quadratically in the Hamiltonian as can easily be seen from (\ref{Ham_second_order}), and (c) proportional to $\sqrt{\Lambda}$ as can also be seen from (\ref{Ham_second_order}), since the $\Lambda$ terms do not have a $P_a$ in front of them.

Once we get this (highly complicated) physical Hamiltonian, the modes can be quantized, and vacuum energy density can be calculated explicitly. It is clear that none of the features mentioned above will change due to quantization. 

We now turn to a model where we can explicitly calculate the vacuum energy density.

\subsection{A toy model}

We now repeat the inhomogeneous mode calculation done in \cite{PhysRevLett.116.061302}, but in the scale factor time gauge (as opposed to volume time).

Our starting point is the full scalar field Hamiltonian given in (\ref{phi_ham}) evaluated on a spatially flat FRW background,
\be
H_{\phi} = \dfrac{\PP^2}{2 a^3} + \frac{1}{2} a^3 m^2 \phi^2 + \frac{1}{2} a e^{ab} \partial_a \phi \partial_b \phi
\ee
where $e_{ab}$ is the flat metric. The scalar field and its conjugate momentum are decomposed in modes as,
\bea
\phi(\textbf{x},t) = \int \dfrac{d^3k}{(2\pi)^3} \phi_{\kk}(t) e^{i \kk.\textbf{x}}, \nn \\
\PP(\textbf{x},t) = \int \dfrac{d^3k}{(2\pi)^3} P_{\kk}(t) e^{i \kk.\textbf{x}}.
\eea

The scalar field Hamiltonian then becomes,
\be
H_{\phi} = \int \dfrac{d^3k}{(2\pi)^3} \dfrac{1}{2} \Bigg( \dfrac{P_{\kk}^2}{a^3} + a^3 m^2 \phi_{\kk}^2 + a |{\kk}|^2 \phi_{\kk}^2 \Bigg).
\ee

We now \emph{define} our theory, with the variables $(a,P_a)$ and $(\phi_{\kk},P_{\kk})$ to be,
\be
S = V_0 \int dt \Big( P_a \dot{a} + \int \dfrac{d^3k}{(2\pi)^3} \Big[P_{\kk} \dot{\phi}_{\kk} - NH \Big] \Big),
\ee
where,
\be
H = -\dfrac{P_a^2}{24 a} + a^3 \Lambda + H_{\phi},
\ee
is the Hamiltonian constraint.

Now, we fix the scale factor time gauge, $t = V_0^{1/3} a$. Solving the Hamiltonian constraint for its conjugate momentum, and defining scale invariant field momentum $p_{\kk} = V_0 P_{\kk}$ and scale invariant wave vector $\bar{{\kk}} = V_0^{1/3} {\kk}$, we get the physical Hamiltonian,
\be
H_p = \sqrt{24} t^2 \sqrt{ \Lambda + \int \dfrac{d^3k}{(2\pi)^3} \Bigg[ \dfrac{p_{\kk}^2}{2t^6} + \dfrac{1}{2} \Bigg( \dfrac{|\bar{\kk}|^2}{t^2} + m^2 \Bigg) \phi_{\kk}^2 \Bigg] }.
\ee 

This can be quantized to get the Energy eigenvalues,
\be
E_n = \sqrt{24}t^2 \sqrt{ \Lambda + \int \dfrac{d^3k}{(2\pi)^3} \Big( n + \frac{1}{2} \Big) \omega_{\kk}(t) },
\ee
where,
\be
\omega_{\kk}(t) = \frac{1}{t^3} \sqrt{ \frac{|\bar{\kk}|^2}{t^2} + m^2 }.
\ee

For vacuum energy, we choose $n=0$, and to simplify further, we set $m=0$. The $k$ integral is done upto some high energy cutoff $\bar{K}$. We finally get the vacuum energy density,
\be
\rho_{\mbox{vac}} = \dfrac{E_0}{t^3} = \dfrac{1}{t} \sqrt{ 24 \Bigg( \Lambda + \dfrac{\bar{K}^4}{16 \pi^2 t^4} \Bigg) }.
\ee

We can clearly see from this that the vacuum energy density is a non-linear function of $\Lambda$, is explicitly time dependent and the contribution from the scalar field modes is suppressed at late times (large Universe), leaving the $\Lambda$ term to be dominant.\footnote{Although it doesn't really matter if it is suppressed at late times or not since there is no experimental measure of this vacuum energy density of the full gravity-matter system. It is to be viewed as a calculation (or prediction) for a given value of $\Lambda$, cutoff scale $\bar{K}$ and time $t$.} We can compare this result to the vacuum energy density obtained in a different time gauge (volume time) in \cite{PhysRevLett.116.061302} (re-arranging their Eq. 39),
\be
\rho_{\mbox{vac}} = \dfrac{1}{t} \sqrt{ \dfrac{8}{3} \Bigg( \Lambda + \dfrac{\bar{K}^4}{16 \pi^2 t^{4/3}} \Bigg) }.
\ee

We can see that these two are not the same, hence making it clear that vacuum energy density depends on the choice of a global time function. This is not surprising since this is just a manifestation of the problem of time. Currently, there is no experimental measure of these vacuum energy densities of the gravity-matter system, meaning that the differences in their values can also be viewed as predictions for different choices of time functions, ultimately to be decided by experiment.
\vskip 0.5cm
Now that we have shown how the cosmological constant problem does not arise in an inhomogeneous gravity-matter setting, we turn our attention to asking another important question: what happens in the low energy regime? That is, for local experiments (carried out in some `small' space and time intervals), we know that quantum field theory on a flat background works remarkably well. So can we reduce from our fully non-perturbative theory down to QFT on a flat background under some appropriate conditions? And if so, how is the cosmological constant problem to be understood in that context? We turn to this in the next section.

\section{Reduction to quantum mechanics on flat space} \label{reduction} 

In this section, we show a way to obtain the usual homogeneous scalar field Hamiltonian on a flat background starting from the FRW theory classically.\footnote{I am thankful to Dr. Viqar Husain for suggesting this idea and its interpretation.}${}^{,}$\footnote{See also \cite{Smolin:1993ka} for a similar idea, but in a different time gauge, and in the full non-homogeneous quantum theory.} We assume that the spatial curvature is zero ($\kappa$ = 0) for simplicity. Our purpose here will be to expand the physical Hamiltonian obtained from (\ref{cosmo_ham}) in the volume time gauge under suitable conditions. Once we have that expansion, we will see that it turns out to be the usual homogeneous scalar field Hamiltonian on a flat background.

To this end, we start with the Hamiltonian constraint (\ref{cosmo_ham}) and reintroduce Planck units,
\be
\label{hcfrw}
H =  \dfrac{-P_a^2}{24 a M_P^2}  + a^3 \Lambda M_P^2   + \dfrac{\PP^2}{2 a^3} + \dfrac{1}{2} a^3 m^2 \phi^2,
\ee
with the action,
\be
S = V_0 \int dt [ P_a \dot{a} + \PP \dot{\phi} - NH ],
\ee
where $ V_0 = \int d^3x $ is the volume of space.

We now fix the volume time gauge
\be
\label{t_now}
\Big(\dfrac{t}{t_P}\Big) \lla^3 = V_0 a^3,
\ee
here  $\lla = \frac{1}{\sqrt{\Lambda}} \sim 10^{60}l_P$ is the natural length scale defined by $\Lambda$.\footnote{Interestingly, its numerical value is of the same order of magnitude as the diameter of the observed Universe.} It is introduced here for dimensional reasons. We have also assumed that we are measuring time in Planck units ($t_P$). We also define a dimensionless time as $ \bt = \frac{t}{t_P} $.

The momentum conjugate to this choice of time will be our physical Hamiltonian density and is given as,
\be
H_p = -\dfrac{P_a}{3a^2} \Bigg( \dfrac{\lla^3}{V_0 t_P}  \Bigg),
\ee
it can be seen that $\{t,-H_p\} = 1$.

We now solve (\ref{hcfrw}) for this Hamiltonian which gives (after defining the scale invariant momentum $\pp = V_0 \PP$),
\be
\label{sqrt_ham}
H_p = \dfrac{\lla^3}{V_0} \dfrac{M_P}{t_P} \sqrt{\dfrac{8}{3}} \sqrt{\Lambda M_P^2 + \rho_{\phi}},
\ee
where,
\be
\rho_{\phi} = \dfrac{\pp^2}{2 \bt^2 \lla^6} + \dfrac{1}{2} m^2 \phi^2,
\ee
and the gauge fixed action is,
\be
\label{S_comp}
S^{GF} = \int dt [ \pp \dot{\phi} - \bar{H}_p ],
\ee
where $\bar{H}_p = V_0 H_p$ is the physical Hamiltonian (not the density).

We now come to the crucial step for reduction to quantum mechanics on flat space. We expand the above square root Hamiltonian (\ref{sqrt_ham}) in the late time regime ($\bt \rightarrow \infty$) under the assumption that,
\be
\label{reduce1}
\rho_{\phi} \ll \Lambda M_P^2.
\ee

\begin{figure}
\begin{center}
\includegraphics[width=0.75\columnwidth, height = 0.3\textheight]{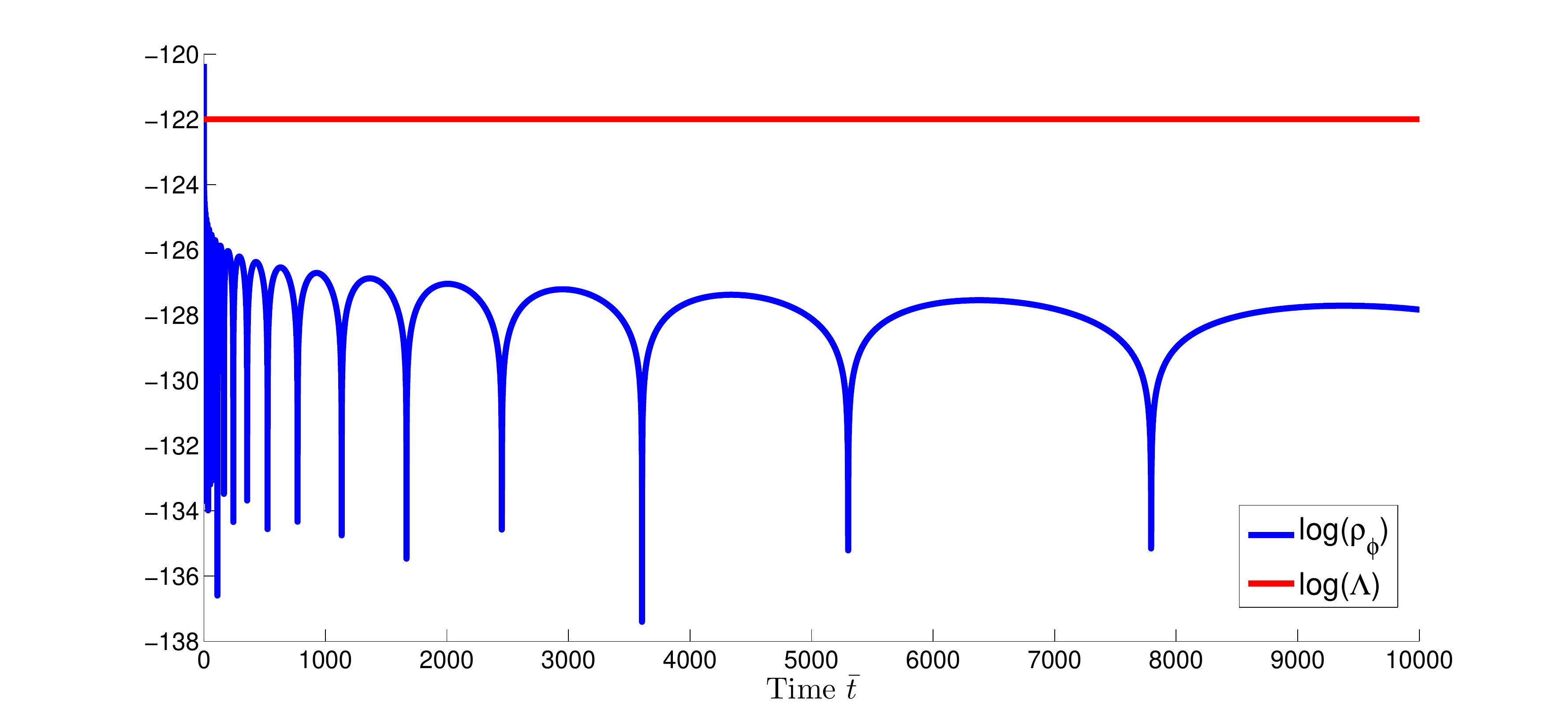}
\end{center}
\caption{log($\rho_{\phi}$) and log($\Lambda$) vs time $\bt$. It can be seen that at late times, $\rho_{\phi} \ll \Lambda$.}
\label{rho_vs_lam}
\end{figure}

The current physical volume of the universe ($V_0 a^3$) is observed to be around $10^{184} l_P^3$, where $l_P$ is the Planck length. Therefore the current time as defined above in (\ref{t_now}) is $\bar{t}_{now} \sim 10^3 $. Hence, the first term in $\rho_{\phi}$ is naturally suppressed at late times. This restricts the other term to be,
\be
\label{reduce2}
m \phi \ll \sqrt{\Lambda} M_P \sim 10^{-61} M_P^2,
\ee
which is quite small (as compared to the Planck scale, i.e $m \phi \sim M_P^2$), but is required for this expansion. Furthermore, we numerically solved the (classical) equations of motion arising from the Hamiltonian (\ref{sqrt_ham}), with parameter values and initial data consistent with the above requirements. Figure \ref{rho_vs_lam} shows the logarithm of $\Lambda$ and the logarithm of $\rho_{\phi}$ versus the dimensionless time $\bar{t}$ for the values $\Lambda = 10^{-122} M_P^2, m = 10^{-60} M_P, \lla = 10^{60} l_P, \phi(0) = 1 M_P, \pp(0) M_P = 10^{-4}$. It is clear from that graph that at late times the condition (\ref{reduce1}) is satisfied. It is also satisfied for other initial data as long as condition (\ref{reduce2}) is met (the mass of the scalar field has to be really small as compared to the Planck mass). Hence we can carry out the expansion.

The expanded Hamiltonian (to leading order) becomes,
\be
\label{Hexp1}
H = \dfrac{\lla^3}{t_P} \sqrt{\dfrac{2}{3\Lambda}} \Bigg[ \dfrac{\pp^2}{2 \bt^2 \lla^6} + \dfrac{1}{2} m^2 \phi^2 \Bigg] + \widetilde{\Lambda},
\ee
here, we have defined $ \widetilde{\Lambda} = \dfrac{ \lla^3 M_P^2}{t_P} \sqrt{\dfrac{8 \Lambda}{3}} $ which is just a constant shift. Note that since we obtained this reduced Hamiltonian starting from a gravity-matter theory, we can no longer add gravity back to it again. Hence, the constant $\widetilde{\Lambda}$ can be ignored \emph{consistently} since it leaves the matter equations of motion invariant and there is no gravity which sees this constant. The shift symmetry problem (as described in Sec \ref{shift_prob}) and the usual cosmological constant problem does not arise here. (To properly include gravity, we have to go back to the (unreduced) non-perturbative gravity-matter theory.)

Now, consider a canonical transformation to new variables,
\bea
\tpp &=& \sqrt{ \dfrac{1}{(\lla \tl)^3}\dfrac{1}{t_P} \sqrt{\dfrac{2}{3 \Lambda}} } ~ \dfrac{\pp}{\bt}, \\
\tphi &=& \sqrt{ \Bigg(\dfrac{\lla}{\tl}\Bigg)^3 t_P \sqrt{\dfrac{3 \Lambda}{2}} } ~ \bt \phi,
\eea
where $\tl$ is an arbitrary length scale. Under this transformation, the Lagrangian becomes,
\bea
\pp \dot{\phi} - H(\phi,\pp,\bt) = \tl^3 \tpp \dot{\tphi} - \widetilde{H}(\tphi,\tpp,\bt), \\ \nn
\widetilde{H}(\tphi,\tpp,\bt) = \tl^3 \Bigg(\dfrac{\tpp^2}{2} + \dfrac{1}{2} \tm^2 \tphi^2 + \dfrac{\tphi \tpp}{t_P \bt} \Bigg),
\eea
where,
\be
\tm = \sqrt{\dfrac{2}{3 \Lambda t_{P}^2}} \dfrac{m}{\bt}. \label{massscale}
\ee

The last term in the Hamiltonian above can be ignored at late times ($\bt \rightarrow \infty$), hence it becomes the usual scalar field Hamiltonian for the `dressed' scalar field $\tphi$. The action for this `dressed' field becomes,
\be
S[\tphi] = \tl^3 \int dt \Bigg( \tpp \dot{\tphi} - \dfrac{\widetilde{H}}{\tl^3} \Bigg),
\ee
which can now be quantized via standard methods. Hence, we have reduced down from the full gravity-matter theory to quantum mechanics on a flat background. 

To find a value for $\tl$ (the box within which this reduced theory is valid), let us consider (\ref{reduce1}) and (\ref{reduce2}) and re-write them in terms of these new variables to see what restrictions they have to satisfy. Eq. (\ref{reduce1}) becomes,
\be
\widetilde{\rho}_{\phi} \ll \sqrt{\dfrac{2 \Lambda}{3}} \Bigg( \dfrac{\lla}{\tl} \Bigg)^3 \dfrac{M_P^2}{t_P},
\ee
where,
\be
\widetilde{\rho}_{\phi} = \dfrac{\tpp^2}{2} + \dfrac{1}{2} \tm^2 \tphi^2.
\ee
Plugging in the current value of $\Lambda$ (\ref{lambda_obs}) and $\lla$ gives,
\be
\widetilde{\rho}_{\phi} \ll \Bigg(\dfrac{3 \times 10^{40} l_P}{\tl}\Bigg)^3  M_P^4.
\ee
We have a choice here: where do we want to impose the cutoff? The smaller the cutoff, the larger the size of the box within which our reduced theory is valid. Assuming that this rescaled scalar field energy density $\widetilde{\rho}_{\phi}$ is much smaller than the Planck scale,
\be
\label{new_rho_scale}
\widetilde{\rho}_{\phi} \ll M_P^4, 
\ee
leads to a value for $\tl$:
\be
\tl \sim 500 ~ \text{km},
\ee
which is 60 times larger than the LHC diameter. Obviously we could also assume cutoff scales much lower than the Planck scale for which this number would increase.\footnote{The highest energy particle observed had an energy of $3 \times 10^{11}$GeV \cite{Bird:1994uy}, 8 orders of magnitude lower than the Planck scale. Assuming this cutoff would give $\tl \sim 2 \times 10^5 ~ \text{km} \sim 1/2$ Earth-Moon distance.}

The other equation (\ref{reduce2}) becomes,
\be
\tm \tphi \ll \sqrt{ \dfrac{M_P^2}{t_P} \sqrt{\dfrac{2 \Lambda}{3}} \Bigg(\dfrac{\lla}{\tl} \Bigg)^3 }.
\ee
Plugging in the above value for $\tl$ then gives
\be
\label{new_phi_scale}
\tm \tphi \ll M_P^2,
\ee
which is quite reasonable as compared to (\ref{reduce2}).

Another interesting thing to note is that the rescaling (\ref{massscale}) makes the low energy mass ($\tm$) time dependent. By similar arguments, any coupling constant would become time dependent through this procedure of reduction. We hope to report more on this in the future \cite{Hassan:vcpaper}.

To summarize, we started with a fundamental `high-energy' theory with scalar field $\phi$, its momentum $\PP$, and mass $m$. The restrictions (\ref{reduce1}) and (\ref{reduce2}), were on the original fundamental degrees of freedom. Once we went through the process of reduction, we found that our theory became that of a usual scalar field, but now for a different field $\tphi$ (dressed in terms of the original field $\phi$). Now, if we quantize this reduced theory, and say that it is the usual quantum mechanics of a scalar field that we observe, then we must make the identification that this new dressed field $\tphi$ is actually the field that we observe (in this reduction regime, not at all times). And that its mass is given by (\ref{massscale}), which changes with time. Therefore, the new restrictions (\ref{new_rho_scale}) and (\ref{new_phi_scale}), are on the field $\tphi$ that we see today, and in comparison to the restrictions on the old field $\phi$, these encompass a larger range of acceptable values (for instance the energy density of the dressed scalar field $\tphi$ can be on the TeV scale, or even beyond). Therefore, we recover usual quantum mechanics of a scalar field, within an acceptable range of validity.

\section{Conclusions} \label{conclusions}

In this section, we list the main conclusions of this work, followed by a summary and discussion in the next section.

\begin{enumerate}

\item The vacuum in a fully non-perturbative quantum theory of gravity and matter is the vacuum of the full system defined as the ground state of the physical Hamiltonian, and depends on the choice of a time function.

\item In general, it is not possible to define a `matter-vacuum', and hence the question: `Does matter-vacuum gravitate?' cannot be framed.

\item In the full theory, at a classical level, the functional form of energy density depends on the chosen time-gauge, and in general, is time dependent and a non-linear (usually square root) function of the cosmological constant. This is to be viewed as a calculation for the energy density, given some observed value of the cosmological constant.

\item There are two special choices of time for which the energy density is linearly proportional to $\Lambda$ (in \emph{contrast} to the results of \cite{PhysRevLett.116.061302}): Dust time, and York time. In York time, it is in fact a spacetime constant.

\item A full quantization is possible (in some time gauges) after a symmetry reduction to cosmology. The vacuum energy density turns out to be time-dependent and proportional to $\sqrt{\Lambda}$. Again, this vacuum energy density is a calculation for a given observed value of $\Lambda$. Moreover, within the cosmological sector, introducing a non-zero spatial curvature leads to a restricted domain for the time function, and performing a Fermionic quantization leads to (almost) the same vacuum energy density as usual Schrodinger quantization.

\item Within a full gravity-matter formalism, a mode expansion is not possible on a dynamical quantum background, and not very useful on a general fixed background due to the Hamiltonian being a square root. In an FRW based toy model where it can be explicitly performed, the cosmological constant problem does not arise since the vacuum energy density is a function of the high energy cutoff, and some \emph{observed} value of the cosmological constant. Furthermore, it depends on the chosen time-gauge, is explicitly time-dependent, and is suppressed at late times (large Universe).

\item There is a way to reduce to quantum mechanics on flat space from the full gravity matter theory in the volume time gauge.

\end{enumerate}

\section{Summary and Discussion} \label{discussion}

The cosmological constant problem as currently understood, comes from a very poor mismatch between theoretical predictions of fixed background QFT and experimental observations of General Relativity. Starting with a fixed background theory and subsequently incorporating gravitational effects is not a sensible thing to do. All of the quantum versions of the problem stated in Sec \ref{cc_prob} are arguments coming from QFT on a fixed (usually flat) background. In this paper, we analyzed the problem in a non perturbative framework.  We have shown that this problem has to be approached non-perturbatively in a fully dynamical theory containing both gravity and matter.

First and foremost, within a fully non-perturbative quantum theory of gravity and matter, a notion of time is \emph{needed} to define what a vacuum is. The understanding of vacuum as the ground state of some Hamiltonian first requires the existence of that Hamiltonian. But in a constrained theory (such as GR), the Hamiltonian vanishes (this is the problem of time), and hence (as shown in Sec \ref{what_vac}) a natural vacuum state \emph{cannot} be defined. One way to proceed (as done in this paper) is to identify a time function on the classical phase space, and then solve (strongly) the Hamiltonian constraint for the variable conjugate to this choice of time, to get a non-vanishing physical Hamiltonian. As we saw above, this depends on the choice of time function, and hence, the ground state, or vacuum of this physical Hamiltonian will depend on the choice of time.

Once we have this physical Hamiltonian and its ground state, vacuum energy density can be calculated by taking the expectation value of this Hamiltonian in the ground state and then dividing by the physical volume. As we saw above, this vacuum energy density, in general, turns out to be time dependent, and is a non-linear function of the cosmological constant. Its exact functional form also depends on the choice of time being made. This vacuum energy density is understood as a \emph{calculation} for some \emph{observed} value of $\Lambda$. 

The conventional cosmological constant problem arises by assuming that quantum matter in its vacuum state will backreact on the background geometry. In a non-perturbative framework however, we have a full gravity-matter system in which both matter and gravity are dynamical, and a matter-vacuum state does not exist in general. Therefore, this question of backreaction is ill-formulated. What is relevant is the vacuum of the full non-pertubative gravity-matter system, which as we saw above, is calculated using some value of $\Lambda$. Therefore, within this non-perturbative framework, the cosmological constant problem does not arise. There is no backreaction since we are already considering the full gravity-matter system. We would like to emphasize that this is a very conventional approach in which we neither modified General Relativity, nor tampered with quantum theory. We defined vacuum as in usual quantum mechanics and then calculated the vacuum energy density in the presence of a cosmological constant term. 

This raises some questions: (i) What is the meaning of these various vacuum energy densities? (ii) Is there any relation between them? (iii) Do they have any observational consequences? and, (iv) Do they give some criteria to judge which time choice is a good one?

Firstly, these vacuum energy densities are not to be confused with the energy densities appearing in the Friedmann equations, these are different. These are the vacuum energy densities of the fully non-perturbative gravity-matter system. They are to be thought of as the proper variables of interest with regards to the cosmological constant problem. Furthermore, each of these vacuum energy densities is with respect to some physical clock (some phase space variable), and hence constitutes a well-defined, gauge-invariant calculation: what is the value of vacuum energy density when a chosen phase space variable (eg scalar field, dust field, spatial volume, etc), which is identified as time, takes a particular value. It is expected that the answer would depend, in general, on what phase space variable is chosen as time, and this was explicitly demonstrated here.

Secondly, it is not clear if there is a relation between them (that could transform one to the other), or if they are fundamentally distinct. This issue is directly related to the problem of time, and we have not solved the problem of time here. It is the same question as: which of all the different Hamiltonians is the `true' Hamiltonian? This cannot be answered until the problem of time is solved. Some possible avenues of exploration (which lie beyond the scope of this paper) are to consider the quantum theories with different choices of time (i.e with different Hamiltonians) and see if: (a) They reproduce QFT on a flat spacetime in some appropriate limit, and hence reproduce all known observations (a hint of this was provided in Sec (\ref{reduction}). See also \cite{Smolin:1993ka} for an answer to this in the full quantum theory). (b) They predict the emergence of an FRW universe at `late times' and (c) They predict some new phenomenon or behavior that could be experimentally detected.

We would also like to mention, that within the covariant formalism as well, energy density \emph{is observer dependent}. An observer with a four velocity $u^{\mu}$, sees an energy density $\rho = T_{\mu \nu} u^{\mu} u^{\nu}$, whereas a different observer with a different four velocity $\tilde{w}^{\mu}$ sees, in general a \emph{different} energy density $\tilde{\rho} = T_{\mu \nu} \tilde{w}^{\mu} \tilde{w}^{\nu} \neq \rho$. The various time slicings performed here can also be thought of as different observers traveling on different worldlines.

Thirdly, currently there is no experimental measure of these vacuum energy densities of the full gravity-matter system. What is experimentally observed is the value of the cosmological constant, obtained after a fit to the FRW universe. Therefore, at this stage, until a method is developed to measure these quantities experimentally, they can be thought of as calculations or predictions for different choices of time, given an experimentally observed value of $\Lambda$. In case it turns out that this quantity can not be measured experimentally, then methods (a-c) listed above (and possibly more) would be used to decide which Hamiltonian, and hence which energy density is the ``right'' one, and then it will remain a calculation, again based on some observed value of $\Lambda$.

Finally, the calculation of energy densities with various choices of time shows how its functional form changes by changing the time gauge, and makes it clear that they can be drastically different from one time choice to another. Moreover, it also gives an indication of which time choices are better than others. This is because we want the resultant physical Hamiltonian to be real. Introducing this constraint then restricts the domain of the time function. Ideally we want our time function to be a globally valid choice. In this context then, dust time is a good choice since it introduces no constraints on the time variable just to keep the Hamiltonian real.

Given our current situation in an expanding universe, the choice of using volume of the universe as a clock seems very natural. We found that with this choice of time, there is a way to reduce from the non-perturbative gravity matter theory to quantum mechanics on a flat background. Starting classically in the volume time gauge along with some assumptions leads to the recovery of the usual homogeneous scalar field Hamiltonian (in a box) which can then be quantized via standard methods. This reduction suggests that volume time is also a good time choice.

Some of the questions that still remain unanswered in this framework are: (i) What is a good time? In our current context, it looks like volume time is very reasonable. Dust time, in which the Hamiltonian is not a square root is also a good candidate. (ii) Related to the first one is the general issue of the problem of time in quantum gravity. Choosing a different time function leads to a physically different Hamiltonian and hence a different quantum theory. Is there a natural choice? (iii) Is there a general method to reduce to flat space QFT starting from a non-perturbative theory? (iv) Why is $\Lambda$ what it is? In our framework, we did not answer the naturalness issue relating to $\Lambda$. We left it to be determined by observations, as parameters in a theory usually are.

This approach (developed in \cite{PhysRevLett.116.061302} and extended here) provides a look beyond flat space QFT arguments. The relation  between vacuum energy density and the cosmological constant is defined consistently only in a holistic approach, one where both gravity and matter are present along with a choice of time and a physical Hamiltonian. Within this framework then, there is no cosmological constant problem.

\begin{acknowledgments}
I am very thankful to Dr. Viqar Husain for numerous discussions, suggestions, motivation and useful comments on the manuscript without which this work would not have been possible. I also thank Masooma Ali, Dr. Jack Gegenberg, Dr. Sanjeev Seahra and Dr. Jon Ziprick for providing useful comments on various versions of the manuscript. I would also like to thank anonymous referees whose comments and questions led to significant improvements in the manuscript. I was supported by the Lewis Doctoral Fellowship.
\end{acknowledgments}

\bibliography{Universe_CC_bib}

\end{document}